\documentclass[amsmath,amssymb,aps,prd,11pt,tightenlines,superscriptaddress,
nofootinbib,preprintnumbers,notitlepage]{revtex4-1}

\usepackage[T1]{fontenc}
\usepackage[utf8]{inputenc}
\usepackage[british]{babel}

\usepackage{color,xcolor}
\definecolor{dark-red}{rgb}{0.8, 0.0, 0.1803921568627451}
\definecolor{dark-blue}{rgb}{0.0, 0.0, 0.803921568627451}
\definecolor{dark-green}{rgb}{0.0, 0.39215686274509803, 0.0}
\definecolor{dark-orange}{rgb}{0.8, 0.4, 0.0}

\usepackage{amsmath,dsfont,mathtools}
\usepackage{siunitx}
\usepackage{graphicx}
\usepackage{multirow,booktabs,tabularx,float}
\usepackage{slashed,braket}
\usepackage{feynmp-auto}
\usepackage[sort&compress]{natbib}
\PassOptionsToPackage{hyphens}{url}\usepackage[colorlinks=true, urlcolor=dark-blue, anchorcolor=dark-blue, citecolor=dark-blue, filecolor=dark-blue, linkcolor=dark-blue, menucolor=dark-blue, pagecolor=dark-blue, linktocpage=true, pdfa=true, linktoc=all]{hyperref}
\usepackage{enumitem} 
\usepackage[stretch=15,shrink=15]{microtype} 
\usepackage{tabto}
\usepackage{accents} 
\usepackage{tikz}
\usepackage{xspace} 
\usetikzlibrary{arrows, decorations.markings, arrows.meta}
\usepackage{pythonhighlight} 

\hypersetup{pdfpagemode=UseNone}

\makeatletter
\def\l@subsubsection#1#2{}
\makeatother

\AtBeginDocument{
  \heavyrulewidth=.08em
  \lightrulewidth=.05em
  \cmidrulewidth=.03em
  \belowrulesep=.65ex
  \belowbottomsep=0pt
  \aboverulesep=.4ex
  \abovetopsep=0pt
  \cmidrulesep=\doublerulesep
  \cmidrulekern=.5em
  \defaultaddspace= .5em
  \setlength{\tabcolsep}{0.5em}
}

\newcommand{\madminer}{\texttt{MadMiner}\xspace}
\newcommand{\toolfont}[1]{\texttt{#1}}

\setlist[itemize]{itemsep=1pt,parsep=1pt, topsep=1pt}

\newcommand{\diff}{\mathrm{d}}
\newcommand{\eg}{{e.\,g.}~}
\newcommand{\ie}{{i.\,e.}~}

\newcommand{\ope}[1]{\mathcal{O}_{#1}}

\newcommand{\met}{{\ensuremath{E_T^{\text{miss}}}}}

\newcommand{\gev}{\si{GeV}}
\newcommand{\tev}{\si{TeV}}

\newcommand{\iab}{\si{ab^{-1}}}

\newcommand{\equref}[1]{Eq.\;\eqref{eq:#1}}

\newcommand{\secref}[1]{Sec.~\ref{sec:#1}}

\newcommand{\figref}[1]{Fig.~\ref{fig:#1}}

\newcommand{\nde}{\textsc{NDE}\xspace}
\newcommand{\alice}{\textsc{Alice}\xspace}
\newcommand{\scandal}{\textsc{Scandal}\xspace}
\newcommand{\alices}{\textsc{Alices}\xspace}
\newcommand{\carl}{\textsc{Carl}\xspace}
\newcommand{\rolr}{\textsc{Rolr}\xspace}
\newcommand{\sally}{\textsc{Sally}\xspace}
\newcommand{\sallino}{\textsc{Sallino}\xspace}
\newcommand{\cascal}{\textsc{Cascal}\xspace}
\newcommand{\rascal}{\textsc{Rascal}\xspace}

\newcommand{\thetaref}{\theta_{\text{ref}}}

\newcommand{\intx}{\int \! \diff x\;}

\newcommand{\threematr}[9]{\begin{pmatrix*}[r] #1 & #2 & #3\\ #4 & #5 & #6 \\ #7 & #8 & #9\end{pmatrix*}}

\newlength{\hhatheight}

\DeclareMathOperator{\diag}{diag}
\DeclareMathOperator{\Real}{Re}
\DeclareMathOperator{\Pois}{Pois}

\DeclareMathOperator{\nn}{NN}

\newcolumntype{R}{>{\raggedleft\arraybackslash}X}%
\newcolumntype{L}{>{\raggedright\arraybackslash}X}%

\newcommand{\be}{\begin{equation}}
\newcommand{\ee}{\end{equation}}

\newcommand{\maa}{m_{\gamma\gamma}}
\newcommand{\ptaa}{p_{T,\gamma\gamma}}

\begin{document}

\count\footins = 1000 

\title{MadMiner: Machine learning--based inference for particle physics}

\preprint{UCI-TR-2019-16, SLAC-PUB-17461}

\author{Johann Brehmer}
\email{johann.brehmer@nyu.edu}
\affiliation{Center for Data Science and Center for Cosmology and Particle Physics,
New York University, New York, NY 10003, USA}

\author{Felix Kling}
\email{felixk@slac.stanford.edu}
\affiliation{Department of Physics and Astronomy, University of
California, Irvine, CA 92697, USA}
\affiliation{SLAC National Accelerator Laboratory, 2575 Sand Hill Road, Menlo Park, CA 94025, USA}

\author{Irina Espejo}
\email{iem244@nyu.edu}
\affiliation{Center for Data Science and Center for Cosmology and Particle Physics,
New York University, New York, NY 10003, USA}

\author{Kyle Cranmer}
\email{kyle.cranmer@nyu.edu}
\affiliation{Center for Data Science and Center for Cosmology and Particle Physics,
New York University, New York, NY 10003, USA}

\begin{abstract}
  Precision measurements at the LHC often require analyzing high-dimensional event data for subtle kinematic signatures, which is challenging for established analysis methods. Recently, a powerful family of multivariate inference techniques that leverage both matrix element information and machine learning has been developed. This approach neither requires the reduction of high-dimensional data to summary statistics nor any simplifications to the underlying physics or detector response. In this paper we introduce \madminer{}, a Python module that streamlines the steps involved in this procedure. Wrapping around \toolfont{MadGraph5\_aMC} and \toolfont{Pythia~8}, it supports almost any physics process and model. To aid phenomenological studies, the tool also wraps around \toolfont{Delphes~3}, though it is extendable to a full \toolfont{Geant4}-based detector simulation. We demonstrate the use of \madminer{} in an example analysis of dimension-six operators in $ttH$ production, finding that the new techniques substantially increase the sensitivity to new physics.
\end{abstract}

\maketitle
\tableofcontents

\section{Introduction}

Precision measurements at the Large Hadron Collider (LHC) experiments
search for direct and indirect signals of physics beyond the Standard Model.
Statistically, this requires constraining a typically high-dimensional parameter
space, for instance the Wilson coefficients in an effective field theory (EFT)
or the couplings and masses in a supersymmetric model. The data going into these
analyses consists of a large number of observables, many of which can carry
information on the parameters of interest.

The relation between model parameters and observables is typically best
described by a suite of computer simulation tools for the hard interaction,
parton shower, hadronization, and detector response. These tools take as input
assumed parameters of the physics model, for instance a particular value for the
Wilson coefficients of an EFT, and use Monte-Carlo methods to sample
hypothetical observations. Unfortunately, they do not directly let us solve the
inverse problem: given a set of observed events, it is not possible to
explicitly calculate the likelihood of such a measurement as a function of the
theory parameters. This intractability of the likelihood function is a major
challenge for particle physics measurements.

Particle physicists have developed a range of techniques for this problem of
\emph{likelihood-free inference}. These can be roughly grouped into three
categories~\cite{Brehmer:2019bvj}:
\begin{enumerate}
 \item{} Traditionally, analyses are restricted to a small number of hand-picked
 observables. The likelihood function for these low-dimensional summary
 statistics can then be estimated with explicit parametric functions,
 histograms, kernel density estimation techniques, or Gaussian
 Processes~\cite{Cranmer:2000du, Cranmer:2012sba, Frate:2017mai}. Relatedly,
 Approximate Bayesian Computation~\cite{rubin1984, beaumont2002approximate,
 Alsing:2018eau, Charnock:2018ogm} is a family of Bayesian techniques that allow
 sampling from an approximate version of the posterior in the space of the
 summary statistics. Coming up with the newest and greatest kinematic
 observables is a popular pastime among phenomenologists. But limiting the
 analysis to a few summary statistics discards the information in all other
 directions in phase space. Even well-motivated variables often do not come
 close to the power of an analysis of the fully differential cross
 section~\cite{Brehmer:2016nyr, Brehmer:2017lrt}.
\item{} Another approach aims to estimate the likelihood function of
high-dimensional observables by approximating the effect of shower,
hadronization, and detector response with simple transfer functions (or
neglecting them altogether). In this approximation, the likelihood becomes
tractable. This category includes the Matrix Element Method~\cite{Kondo:1988yd,
Abazov:2004cs, Artoisenet:2008zz, Gao:2010qx, Alwall:2010cq, Bolognesi:2012mm,
Avery:2012um, Andersen:2012kn, Campbell:2013hz, Artoisenet:2013vfa,
Gainer:2013iya, Schouten:2014yza, Martini:2015fsa, Gritsan:2016hjl,
Martini:2017ydu, Kraus:2019qoq}, Optimal Observables~\cite{Atwood:1991ka,
Davier:1992nw, Diehl:1993br}, and Shower and Event
Deconstruction~\cite{Soper:2011cr, Soper:2012pb, Soper:2014rya,
Englert:2015dlp}. These methods make maximal use of the knowledge about the
physics underlying the simulations. While these methods do not require picking
summary statistics, the approximation of the detector response can lead to
suboptimal results, the treatment of additional jet radiation is a challenge,
and the evaluation of each event requires the calculation of a numerically
expensive integral.
\item{} Over the last years methods based on machine learning have become
increasingly popular. The industry standard in particle physics is to train a
classifier (often a boosted decision tree or neural network) to classify events
as coming from different sources (\eg signal vs.~background). Its output is used
to define acceptance regions, accepted events are then usually analyzed with  a
traditional histogram-based measurement strategy. While this strategy is great
at suppressing background events, it does not necessarily lead to the most
precise parameter measurements when kinematic distributions change over the
parameter space~\cite{Brehmer:2016nyr}.

Only recently has there been an increased interest in using machine learning to
estimate the likelihood, likelihood ratio, or (in a Bayesian setting) the
posterior~\cite{2012arXiv1212.1479F,
2014arXiv1410.8516D, 2015arXiv150203509G,
Cranmer:2015bka, Cranmer:2016lzt, Louppe:2016aov,
2016arXiv160508803D, 2016arXiv160506376P,
2016arXiv161110242D, 2016arXiv160502226U,
gutmann2017likelihood, 2017arXiv170208896T, 2017arXiv170707113L,
2017arXiv170507057P, 2017arXiv171101861L, 2018arXiv180400779H,
2018arXiv180507226P, 2018arXiv180509294L, DBLP:journals/corr/abs-1806-07366,
2018arXiv180703039K,  2018arXiv181001367G, 2018arXiv181009899D, Hermans:2019ioj,
Alsing:2019xrx, 2019arXiv190507488G}. These approaches have in common that they only require access
to samples generated for different model parameter values. They can handle
high-dimensional observables and do not require a choice of summary statistics.
They also work natively with the output of the simulator, so they do not require
any simplifications to the underlying physics or detector response. The estimate
of the likelihood provided by these algorithms typically becomes exact in the
limit of infinite training samples (assuming sufficient capacity and efficient
training), but often a large number of simulations is required before a good
performance is reached.
\end{enumerate}

A new machine-learning-based approach that directly leverages matrix element
information has been introduced in Refs.~\cite{Brehmer:2018hga, Brehmer:2018kdj,
Brehmer:2018eca} and since been further developed in Refs.~\cite{Stoye:2018ovl,
Brehmer:2019bvj}. Like the other multivariate approaches, these techniques
support high-dimensional observables without the restriction to summary
statistics. Similar to the Matrix Element Method and Optimal Observables, these
techniques leverage our physics insight in the form of the matrix elements
efficiently. But unlike those methods, they support state-of-the-art simulations
of the parton shower and detector response. In addition, after an upfront
simulation and training phase, they provide a function that estimates the
likelihood and can be evaluated in microseconds.

These new techniques require extracting matrix-element information from the
Monte-Carlo simulation, keeping track of and manipulating these weights in
specific ways, and then training machine learning models on this data. Without
proper software support, these steps are cumbersome and error-prone, providing a
technological hurdle to a wider adaptation of these methods.
Reference~\cite{Brehmer:2018hga} describes this approach with the analogy of
``mining gold'' from Monte-Carlo simulations: while the additional information
from the simulations is very valuable, it can require some effort to extract and
process. But the gold does not have to be hard to mine!

In this paper we introduce \madminer{}, a Python module that automates all
steps necessary for these modern multivariate inference techniques. It supports
all elements of a typical analysis, including the simulation of events with
\toolfont{MadGraph5\_aMC}~\cite{Alwall:2014hca},
\toolfont{Pythia~8}~\cite{Sjostrand:2007gs}, detector simulation, reducible and
irreducible backgrounds, and systematic uncertainties. For phenomenological
studies, the tool supports the simulation of the detector response with
\toolfont{Delphes~3}~\cite{deFavereau:2013fsa}, though it is extendable to a
full detector simulation based on \toolfont{Geant4}~\cite{Agostinelli:2002hh}.

We review the supported analysis techniques in Sec.~\ref{sec:inference} and
describe their implementation in \madminer{} in Sec.~\ref{sec:implementation}.
In Sec.~\ref{sec:example}, the new tool is demonstrated in an example analysis
of Higgs production in association with a top pair at the high luminosity run of
the LHC. We give our conclusions in Sec.~\ref{sec:conclusions}. In the appendix
we answer frequently asked questions.

\section{Inference techniques}
\label{sec:inference}

\subsection{LHC measurements as a likelihood-free inference problem}
\label{sec:likelihood-free}

The ultimate goals of most measurements are best-fit points and exclusion
regions in a (high-dimensional) parameter space. In particle physics
experiments, best-fit points are typically defined as maximum likelihood
estimators, while exclusion regions are based on hypothesis tests that use the
(profile) likelihood ratio as test
statistic~\cite{Cranmer:2015nia}.\footnote{The issue of likelihood-free
inference, the inference techniques discussed here, and \madminer{} just as well
apply in a Bayesian setting, see for instance Ref.~\cite{Hermans:2019ioj}.} Both
are based on the same central object, the likelihood function
$p_\text{full}(\{x\}|\theta)$. It quantifies the probability of observing a
set of events, where each event is characterized by a vector $x$ of
observables such as reconstructed energies, momenta, and angles of all
final-state particles, as a function of a vector of model parameters $\theta$,
\eg the Wilson coefficients of an effective field theory.

In particle physics measurements, the likelihood function usually has the form
\be
  p_\text{full}(\{x\}|\theta) = \Pois(n | L \sigma(\theta)) \; \prod_i p (x_i | \theta) \,.
  \label{eq:extended_likelihood}
\ee
Here $n$ is the observed number of events, $L$ is the integrated luminosity,
$\sigma(\theta)$ is the cross section as a function of the model parameters,
$\Pois(n|\lambda) = \lambda^n e^{-\lambda} / n!$ is the probability mass
function of the Poisson distribution, and
\be
  p(x|\theta) = \frac 1{\sigma(x)}  \frac {\diff^d \sigma(x|\theta)} {\diff x^d}
\ee
is the likelihood function for a single event: the probability density of the
$d$-dimensional vector of observables $x$ as a function of the model parameters
$\theta$. Up to the normalization, this kinematic likelihood function is
identical to the fully differential cross section $\diff^d \sigma(x|\theta) /
\diff x^d$.

The Poisson or rate term in Eq.~\eqref{eq:extended_likelihood} is comparably
simple, even though it is based on the cross section after efficiency and
acceptance effects, which can be complicated to calculate in realistic problems.
But the remaining terms, which quantify the kinematic information, typically
cannot be explicitly computed at all. This is because the most accurate model of
the kinematic distributions is usually given by a complicated chain of
Monte-Carlo simulators. The kinematic likelihood they implicitly define can be
written symbolically as
\be
  p(x|\theta) = \int\!\diff z_d \; \int\!\diff z_s \; \int\!\diff z_p \;
  \underbrace{p(x|z_d) \, p(z_d|z_s) \, p(z_s|z_p) \, p(z_p|\theta)}_{p(x,z|\theta)} \,,
  \label{eq:intractable_integral}
\ee
where $z_p$ are the four-momenta, charges, and helicities of the parton-level
four-momenta, $z_s$ is the entire history of the parton shower, and $z_d$
describe the interactions of the particles with the detector. A state-of-the-art
simulation can easily involve billions of such latent variables. Explicitly
calculating the integral over this huge space is clearly impossible: given a set
of events $\{x\}$ and a parameter point $\theta$, we hence cannot compute the
likelihood function (it is \emph{intractable}). This is a major challenge for
analyzing LHC data. The same structural problem appears in many other fields
that use computer simulations to model complicated processes, including
cosmology, systems biology, and epidemiology, giving rise to the development of
different likelihood-free inference techniques.

In particle physics, common analysis techniques address the intractability of
the likelihood function in different ways. The traditional approach restricts
the observables $x$ to one or two summary statistics $v(x)$, for instance the
invariant mass of the decay products of a searched resonance or the transverse
momentum of the hardest particle in an EFT analysis. Then the density
$p(v|\theta)$ can be calculated with histograms and used in lieu of the full
likelihood $p(x|\theta)$. On the other hand, the Matrix Element Method and
Optimal Observable approaches simplify the integral in
Eq.~\eqref{eq:intractable_integral} by replacing the shower and detector
response with simple smearing or transfer functions; in this approximation it
also becomes tractable. For a discussion and comparison of these different
methods see Ref.~\cite{Brehmer:2019bvj}.

\subsection{Learning the likelihood function}
\label{sec:likelihood_estimators}

A first class of inference techniques in \madminer{} tackles the intractability
of the likelihood function head-on: a neural network is trained to estimate the
kinematic likelihood $p(x|\theta)$ or, equivalently, the likelihood ratio
\be
  r(x|\theta_0, \theta_1) = \frac {p(x|\theta_0)} {p(x|\theta_1)}
\ee
using data available from the simulator. To be more specific, \madminer{}
differentiates between three different functions that the neural network can
learn:
\begin{description}
  \item[Likelihood estimators] In this case, a neural network takes as input
  event data $x$ as well as a model parameter point $\theta$ and returns the
  estimated likelihood $\hat{p}(x | \theta)$,
    \be
    \nn: (x, \theta) \mapsto \hat{p}(x | \theta) \,.
    \ee

  \item[Likelihood ratio estimators]
  Alternatively, the
  network can model the likelihood ratio including its dependence on the data
  $x$ and on the parameter point $\theta$ in the numerator of the ratio,
    \be
    \nn: (x, \theta) \mapsto \hat{r} (x | \theta) \approx \frac {p(x | \theta)} {p_{\text{ref}}(x)} \,.
    \ee
    There are different options for the denominator distribution
    $p_{\text{ref}}(x)$. In \madminer{} we set it to the distribution from a
    reference parameter point, $p_{\text{ref}}(x) = p(x | \thetaref)$.
    Alternatively, it could be given by a marginal model $p_{\text{ref}}(x) =
    \int \! \diff \theta' \; p(x | \theta') p(\theta')$, or even be an entirely
    unphysical reference distribution.

  \item[Doubly parameterized likelihood ratio estimators] The last option is to
  model the likelihood ratio as a function of not only the event data $x$ and
  the numerator parameter point $\theta_0$, but also including its
  dependence on the denominator model,

    \be
    \nn: (x, \theta_0, \theta_1) \mapsto \hat{r}(x | \theta_0, \theta_1)
    \approx \frac {p(x | \theta_0)} {p(x | \theta_1)} \,.
    \ee
\end{description}

Note that in all three cases, the network is \emph{parameterized} in terms of the
theory parameters $\theta$~\cite{Cranmer:2015bka, Baldi:2016fzo}: rather than
training separate networks for different points on a grid of parameter points,
one neural network models the likelihood function for the whole parameter
space. The network learns to interpolate in parameter space and can ``borrow''
statistical power from close parameter points, leading to a significantly better
sample efficiency than a point-by-point approach~\cite{Brehmer:2018eca}.

\begin{table}
\begin{tabular*}{0.95\textwidth}{@{\extracolsep{\fill}} l l c@{~}c c c r}
\toprule
\multirow{2}{*}{Method} & \multirow{2}{*}{Run simulation at} & \multicolumn{2}{c}{Loss fn.~uses} & \multirow{2}{*}{Asympt.~exact} & \multirow{2}{*}{Generative} & \multirow{2}{*}{Ref.} \\[-1pt]
& & $r(x, z)$ & $t(x, z)$ \\
\midrule
\multicolumn{7}{l}{\textbf{Likelihood estimators}} \\
\nde & $\theta \sim \pi(\theta)$ & & & $\checkmark$ & $\checkmark$ & \cite{2017arXiv170507057P} \\
\scandal & $\theta \sim \pi(\theta)$ & & $\checkmark$ & $\checkmark$ & $\checkmark$ & \cite{Brehmer:2018hga} \\
\midrule
\multicolumn{7}{l}{\textbf{Likelihood ratio estimators}} \\
\carl & $\theta \sim \pi(\theta)$, $\thetaref$ & & & $\checkmark$ & & \cite{Cranmer:2015bka} \\
\rolr & $\theta \sim \pi(\theta)$, $\thetaref$ & $\checkmark$ & & $\checkmark$ &   & \cite{Brehmer:2018eca} \\
\alice& $\theta \sim \pi(\theta)$, $\thetaref $ & $\checkmark$ & & $\checkmark$ &  & \cite{Stoye:2018ovl} \\
\cascal& $\theta \sim \pi(\theta)$, $\thetaref$ &  & $\checkmark$ & $\checkmark$ &   & \cite{Brehmer:2018eca} \\
\rascal& $\theta \sim \pi(\theta)$, $\thetaref$ & $\checkmark$ & $\checkmark$ & $\checkmark$ &  & \cite{Brehmer:2018eca} \\
\alices& $\theta \sim \pi(\theta)$, $\thetaref$ & $\checkmark$ & $\checkmark$ & $\checkmark$ &  & \cite{Stoye:2018ovl} \\
\midrule
\multicolumn{7}{l}{\textbf{Doubly parameterized likelihood ratio estimators}} \\
\carl & $\theta_0 \sim \pi(\theta)$, $\theta_1 \sim \pi(\theta)$ & & & $\checkmark$ & & \cite{Cranmer:2015bka} \\
\rolr & $\theta_0 \sim \pi(\theta)$, $\theta_1 \sim \pi(\theta)$ & $\checkmark$ & & $\checkmark$ &  & \cite{Brehmer:2018eca} \\
\alice& $\theta_0 \sim \pi(\theta)$, $\theta_1 \sim \pi(\theta)$ &  $\checkmark$ & & $\checkmark$ & & \cite{Stoye:2018ovl} \\
\cascal& $\theta_0 \sim \pi(\theta)$, $\theta_1 \sim \pi(\theta)$ &  & $\checkmark$ & $\checkmark$ &   & \cite{Brehmer:2018eca} \\
\rascal& $\theta_0 \sim \pi(\theta)$, $\theta_1 \sim \pi(\theta)$ & $\checkmark$ & $\checkmark$ & $\checkmark$ & & \cite{Brehmer:2018eca} \\
\alices& $\theta_0 \sim \pi(\theta)$, $\theta_1 \sim \pi(\theta)$ & $\checkmark$ & $\checkmark$ & $\checkmark$ & & \cite{Stoye:2018ovl}  \\
\midrule
\multicolumn{7}{l}{\textbf{Score estimators}} \\
\sally & $\thetaref$ & & $\checkmark$ & in local approx.& & \cite{Brehmer:2018eca} \\
\sallino & $\thetaref$ & & $\checkmark$ & in local approx.&  & \cite{Brehmer:2018eca} \\
\bottomrule
\end{tabular*}
\caption{Inference techniques implemented in \madminer. We separate them into
four groups, depending on which quantity is estimated by the neural network; see
the text for more details. We give the parameter values for which the Monte-Carlo samples
have to be generated and list whether the augmented data (joint likelihood ratio
$r(x, z)$ and joint score $t(x,z)$) are used. ``Asymptotically exact''
describes methods that should give theoretically optimal results in the
limit of sufficient network capacity, perfect optimization, and enough training
data. Methods that also allow for the fast generation of event data
from the neural network are marked as ``generative''. Finally,
for each method we provide the reference that provides the clearest explanation
(and spells out the acronym).}
\label{tbl:inference_techniques}
\end{table}

But how do we train a neural network to learn any of these three functions? More
specifically, which loss function can we minimize so that a neural network will
converge to the right function? There are a number of different answers, which
can be grouped into two categories. First, some methods have been developed that
just use samples of events $\{x\}$ generated from different parameter points
$\theta$. This includes \emph{neural density estimation} (\nde) techniques, for
instance masked autoregressive flows~\cite{2017arXiv170507057P}, in which the
network learns the likelihood function. Another approach is the \carl
method~\cite{Cranmer:2015bka}, which trains the network to estimate the
likelihood ratio.

While both \nde and \carl are implemented in \madminer{}, its major feature is
the support for a new, potentially more powerful paradigm to likelihood or
likelihood ratio estimation~\cite{Brehmer:2018hga, Brehmer:2018kdj,
Brehmer:2018eca}. The key idea is that additional information can be extracted
from the Monte-Carlo simulations, and that this additional information can be
used to train more precise estimators of likelihood or likelihood ratio with
less training data.

More specifically, for each simulated event it is possible
to calculate the joint likelihood ratio
\be
  r(x,z | \theta_0, \theta_1)
  \equiv \frac {p(x, z | \theta_0)} {p(x, z | \theta_1)}
  = \frac {p(x|z_d) \, p(z_d|z_s) \, p(z_s|z_p) \, p(z_p|\theta_0)}
  {p(x|z_d) \, p(z_d|z_s) \, p(z_s|z_p) \, p(z_p|\theta_1)}
  = \frac {\diff \sigma(z_p | \theta_0)} {\diff \sigma(z_p | \theta_1)} \,
  \frac {\sigma(\theta_1)} {\sigma(\theta_0)}
\ee
and the joint score
\be
  t(x, z | \theta)
  \equiv \nabla_\theta \log p(x, z | \theta)
  = \frac {p(x|z_d) \, p(z_d|z_s) \, p(z_s|z_p) \, \nabla_\theta p(z_p|\theta)}
  {p(x|z_d) \, p(z_d|z_s) \, p(z_s|z_p) \, p(z_p|\theta)}
  = \frac {\nabla_{\theta} \diff \sigma(z_p | \theta)} {\diff \sigma(z_p | \theta)}
  - \frac {\nabla_{\theta} \sigma(\theta)} {\sigma(\theta)}
  \,.
\ee
Here $\sigma(\theta)$ is the total cross section as a function of the model
parameters $\theta$, and $\diff \sigma(z_p | \theta)$ are the parton-level event
weights. At a hadron collider such as the LHC these can be written as~\cite{Alwall:2010cq}
\be
  \diff \sigma(z_p | \theta)
  = \frac {(2\pi)^4 f_1(\mathrm{x}_1, Q^2) f_2(\mathrm{x}_2, Q^2)} {2 \mathrm{x}_1 \mathrm{x}_2 s} \;
   |\mathcal{M}|^2(z_p | \theta) \; \diff \Phi(z_p) \,.
  \label{eq:event_weights}
\ee
They depend on the momentum fractions $\mathrm{x}_i$ carried by the
initial-state partons, the squared center-of-mass energy $s$, the momentum
transfer $Q$, the corresponding values of the parton density functions
$f_i(\mathrm{x}_i, Q^2)$, and the Lorentz-invariant phase-space element $\diff
\Phi(z_p)$. Finally, $z_p$ is the entire phase-space point of a simulated event
(including the parton four-momenta, helicities, and charges), and
$|\mathcal{M}|^2(z_p | \theta)$ is the squared matrix element. Both the joint
likelihood ratio and the joint score thus depend on the parton-level momenta
$z_p$ and are directly related to the squared matrix element describing the
underlying process.

The main insight of Refs.~\cite{Brehmer:2018hga, Brehmer:2018kdj,
Brehmer:2018eca} is that the joint likelihood ratio and joint score can be used
to define loss functions that, when minimized with respect to a test function
that only depends on $x$ and $\theta$, converges to the likelihood function $p(x
| \theta)$ or the likelihood ratio.\footnote{Note that this approach is similar
in spirit to the Matrix Element Method, which also uses parton-level likelihoods
and aims to estimate $r(x | \theta_0, \theta_1)$ by calculating approximate
versions of the integral in \equref{intractable_integral}. But unlike the Matrix
Element Method, our machine-learning-based approach supports realistic shower
and detector simulations and can be evaluated very efficiently.}

There are several variations of this idea. The main difference between them is
the exact form of the loss function used. We label them with a set of acronyms:
\scandal is an improved version of \nde techniques that uses the joint score to
train likelihood estimators more efficiently; \cascal is a similarly improved
version of the \carl method; \rolr and \alice use the joint likelihood ratio to
efficiently train a likelihood ratio estimator; and finally the \rascal and
\alices techniques use both the joint likelihood ratio and the joint score,
maximizing the use of information from the simulator. In
Tbl.~\ref{tbl:inference_techniques} we provide an overview and give references
to detailed explanations of all methods.

Once a neural network has been trained with one of these methods, it can
calculate an estimated value of the likelihood or likelihood ratio for any event
and any parameter point. Established statistical tools can then be used to
calculate best-fit points and exclusion limits in the parameter space. For the
calculation of frequentist confidence regions, there are generally two
strategies. The first is simulating a large number of toy experiments to
calculate the $p$-value for each parameter point that is tested. This approach
can be computationally expensive, but guarantees statistically correct
results\,---\,even if the neural network has not learned the likelihood function
accurately, this approach will not lead to too tight limits. The second strategy
uses the asymptotic properties of the likelihood ratio
function~\cite{Wilks:1938dza, Wald, Cowan:2010js} to directly translate values
of the likelihood ratio into $p$-values. While this method is extremely
efficient, it relies on correctly trained neural networks.

\subsection{Learning locally optimal observables}
\label{sec:score_estimators}

\madminer{} also implements a second class of methods: rather than trying to
reconstruct the full likelihood function, a neural network is trained to provide
the most powerful observables for a given measurement problem. The central
quantity of this approach is the \emph{score}
\be
t(x) = \nabla_\theta \log p(x | \theta) \biggr |_{\thetaref}
\label{eq:score}
\ee
evaluated at a fixed reference parameter point $\thetaref$, for instance the
SM. This vector has one component per parameter. For a given event $x$, its
components are just numbers (unlike the likelihood and the likelihood ratio,
which are also functions of the parameters $\theta$). In other words, the score is a
vector of observables.

The relevance of these observables is most obvious in a local approximation of
the likelihood function~\cite{Alsing:2017var, Brehmer:2018kdj, Alsing:2018eau}:
in the neighborhood of the parameter point $\thetaref$, the score components are
the sufficient statistics. That means that for the purpose of measuring
$\theta$, knowing $t(x)$ is just as powerful as knowing the full likelihood
$p(x|\theta)$ (which, since it depends on $\theta$, is a much more complicated
object). In this sense, the score defines the most powerful observables for the
measurement of $\theta$.\footnote{In fact, the score vector is a generalization
of the concept of Optimal Observables~\cite{Atwood:1991ka, Davier:1992nw,
Diehl:1993br} from the parton level to the full statistical model including
shower and detector simulation.}

This motivates a fourth function for a neural network to estimate:
\begin{description}
  \item[Score estimator] A neural network takes as input event data $x$
  and returns the estimated score at a reference parameter point,
    \be
    \nn: x \mapsto \hat{t}(x) \approx \nabla_\theta \log p(x | \theta) \biggr |_{\thetaref}
    \label{eq:score_estimator}
    \ee
\end{description}

How does a neural network learn to estimate the score? Again, extracting
additional information from the simulator proves useful. The \sally and \sallino
methods introduced in Refs.~\cite{Brehmer:2018hga, Brehmer:2018kdj,
Brehmer:2018eca} define a loss function that involves the joint score $t(x, z)$.
Minimizing this loss function will train a neural network to converge to the true
score $t(x)$~\citep{Brehmer:2018hga}.

After training, such a score estimator can be used like any other set of
observables. In particular, we can fill multivariate histograms of the score and
use them for inference. This approach, named \sally, requires only a minimal
modification of established analysis workflows. A similar method called \sallino
constructs one-dimensional histograms of particular projections of the estimated
score, see Ref.~\cite{Brehmer:2018eca} for details.

As long as parameter points close to the reference point, for instance the SM,
are analyzed, and assuming that the neural network was trained efficiently and
with enough training data, the \sally or \sallino methods will lead to
statistically optimal limits. Further away from the reference point, the score
components might no longer be optimal, and this approach might lose some power
compared to the techniques discussed in the previous section. The size
of the parameter region in which the score components are the sufficient statistics
depends on the size of higher derivatives of the (log) likelihood with
respect to the parameters and is not known a priori; we will illustrate this
with an example in Sec.~\ref{sec:example-results}.

\subsection{The Fisher information}
\label{sec:fisher_info}

The final results of actual measurements are best-fit points and exclusion
limits. However, for quickly evaluating the sensitivity of a measurement,
comparing different channels, or optimizing an analysis, a different object is
often more practical: the \emph{Fisher information} matrix. It is closely
connected to the score discussed in the previous section and summarizes the
sensitivity of an analysis in a compact, interpretable, and powerful
way~\cite{Brehmer:2016nyr, Brehmer:2017lrt}. It is defined as the expectation
value
\be
  I_{ij}(\theta) = \mathbb{E} \left[ \frac {\partial \log p_{\text{full}}(\{x\}|\theta)}{\theta_i} \,
  \frac {\partial \log p_{\text{full}}(\{x\}|\theta)}{\theta_j} \middle| \theta \right]
\ee
with the full likelihood function $p_\text{full}(\{x\}|\theta)$ from
Eq.~\eqref{eq:extended_likelihood}.

To see why this matrix is useful, consider an expansion of the expected log likelihood ratio
between $\theta + \Delta \theta$ and $\theta$ around the minimum:
\begin{align}
-2\; \mathbb{E} \left[\log
\frac {p_{\text{full}}(\{x\}|\theta + \Delta \theta)} {p_{\text{full}}(\{x\}|\theta)}
\middle| \theta \right]
&= - \mathbb{E} \left[
\frac{\partial^2 \log p_{\text{full}}(\{x\}|\theta)} { \partial \theta_i \;\partial \theta_j}
\middle | \theta\right]
\; \Delta \theta_i \; \Delta \theta_j
+ \mathcal{O}(\Delta \theta^3) \notag \\
&= \mathbb{E} \left[
\frac{\partial \log p_{\text{full}}(\{x\}|\theta)} { \partial \theta_i} \;
\frac{\partial \log p_{\text{full}}(\{x\}|\theta)} { \partial \theta_j}
\middle | \theta\right]
\; \Delta \theta_i \; \Delta \theta_j
+ \mathcal{O}(\Delta \theta^3) \notag \\
&=
  I_{ij}(\theta) \; \Delta \theta_i \; \Delta \theta_j
+ \mathcal{O}(\Delta \theta^3) \notag \\
&= d(\theta, \theta + \Delta \theta)^2 + \mathcal{O}(\Delta \theta^3) \,.
\label{eq:fisher_approx}
\end{align}
In the last step we have introduced the \emph{local Fisher distance}
\be
  d(\theta + \Delta \theta, \theta) = \sqrt{I_{ij}(\theta) \; \Delta \theta_i \; \Delta \theta_j}
\ee
which is a convenient approximation of the log likelihood ratio as long as
$\Delta \theta$ is small.\footnote{The Fisher information defines a metric on
the parameter space, giving rise to the field of information
geometry~\cite{efron1975, amari1982, Brehmer:2016nyr}. In that formalism we can
also define ``global'' distances measured along geodesics, which are equivalent
to the expected log likelihood ratio even beyond the local approximation of
small $\Delta \theta$~\cite{Brehmer:2017fyp}.} Moreover, according to the
Cram\'er-Rao bound~\cite{Rao:1945, Cramer:1946} the inverse of the Fisher
information is the minimal covariance of any estimator $\hat{\theta}$. The
larger the Fisher information, the more precisely a parameter can be measured.

This approach shines when it comes to ease of use and interpretability. The
Fisher information matrix is invariant under reparameterizations of the
observables $x$, transforms covariantly under reparameterizations of the
parameters $\theta$, and is additive over phase-space regions. This property
means we can define the distribution of the differential information over phase
space, which quantifies where in phase space the power of an analysis comes
from~\cite{Brehmer:2016nyr}. The formalism also easily accommodates nuisance
parameters, and profiling over them is a simple matrix
operation~\cite{Brehmer:2016nyr, Edwards:2017mnf}.

In particle physics processes described by Eq.~\eqref{eq:extended_likelihood},
the Fisher information turns out to be~\cite{Brehmer:2016nyr}
\begin{align}
I_{ij}(\theta)
&= \frac {L \, \partial_i \sigma(\theta) \, \partial_j \sigma(\theta)} {\sigma(\theta)}
+ L \, \sigma(\theta) \; \intx p(x | \theta) \, t_i(x|\theta) \, t_j(x|\theta) \notag \\
&\approx \frac {L \, \partial_i \sigma(\theta) \, \partial_j \sigma(\theta)} {\sigma(\theta)}
+ \frac {L \, \sigma(\theta)}{n} \; \sum_{x \sim p(x | \theta)}  t_i(x|\theta) \, t_j(x|\theta) \,,
\label{eq:fisher_info_lhc}
\end{align}
where $L$ is the integrated luminosity, $\sigma$ the cross section, $\partial_i$
denotes derivatives with respect to $\theta_i$, $n$ is the number of events
$x$ generated for the parameter point $\theta$, and $t_i$ is the $i$-th component
of the score vector introduced in Eq.~\eqref{eq:score}. The first term describes the
information in the overall rate, while the second term quantifies the power in
the kinematic distributions. A neural score estimator $\hat{t}(x)$ as in
Eq.~\eqref{eq:score_estimator} together with a set of events thus lets us
calculate the (a priori intractable) Fisher information.

\subsection{Practical analysis aspects}
\label{sec:practical_analysis_aspects}

Let us now link these abstract inference techniques to specific aspects of
typical analyses in high-energy physics and summarize some features and
limitations of \madminer{}.

\begin{description}
\item[High-energy process]
\madminer{} supports almost all processes of perturbative high-energy physics
that can be run in \toolfont{MadGraph5\_aMC}~\cite{Alwall:2014hca}. This
includes any high-energy physics model specified in the \toolfont{UFO}
format~\cite{Degrande:2011ua}. The inference techniques only require that the
model is parameterized by a finite number of model parameters $\theta$ and that
it is possible to calculate the parton-level event weights of
Eq.~\eqref{eq:event_weights}  for arbitrary values of the model parameters
$\theta$, \ie to ``reweight'' the events to different parameter
points~\cite{Mattelaer:2016gcx}. The approach is not fundamentally restricted to
leading order, though one has to be careful that negative event weights, which
can appear in certain subtraction schemes, do not lead to numerical
instabilities.

It is often beneficial to define the parameters $\theta$ such that they span a
similar order of magnitude. In practice, this may require some rescaling. For
instance, if an analysis aims to measure two Wilson coefficients $f_0$ and $f_1$
and the range of interest of $f_1$ is 1000 times larger than that of $f_0$, consider
defining the parameters internally as $\theta = (f_0, f_1 / 1000)$.

\item[Morphing]
In an important class of models the squared matrix elements (or parton-level
event weights) can be factorized into a sum over $n_c$ components, each
consisting of an analytical function of the theory parameters times a function of
phase-space:
\be
  |\mathcal{M}|^2(z_p | \theta) = \sum_{c} w_{c}(\theta) \,  f_c(z_p) \,.
  \label{eq:morphing}
\ee
This is often the case in effective field theories, or when indirect effects of
new physics are parameterized through form factors.

For instance, consider the simple case in which we are trying to measure a
single BSM parameter $\theta$ and the process is described by a SM contribution,
an interference term, and a squared BSM amplitude:
\be
  |\mathcal{M}|^2(z_p | \theta)
    = \underbrace{1\vphantom{M_{M}^\dagger}}_{w_0(\theta)\vphantom{f_0}} \, \underbrace{|\mathcal{M}_{SM}|^2(z_p)}_{f_0(z_p)}
    + \underbrace{\theta\vphantom{M_{M}^\dagger}}_{w_1(\theta)\vphantom{f_0}} \,
    \underbrace{2 \Real \mathcal{M}_{SM}^\dagger(z_p) \, \mathcal{M}_{BSM}(z_p)}_{f_1(z_p)}
    + \underbrace{\theta^2\vphantom{M_{M}^\dagger}}_{w_2(\theta)\vphantom{f_0}} \,
    \underbrace{|\mathcal{M}_{BSM}|^2(z_p)}_{f_2(z_p)} \,.
\ee
More generally, the dependency on the model parameters is often a combination of
different polynomials. Note that the contributions $f_c(z_p)$ are not
necessarily distributions: they can be negative, or integrate to zero, for
instance for interference terms. Nevertheless, the sum of all components is
always a physical distribution, \ie it is non-negative everywhere and integrates
to the total cross section.

When a process factorizes according to \equref{morphing}, a ``morphing
technique''~\cite{ATLAS:morphing, Brehmer:2018eca} allows us to calculate event
weights anywhere in parameter space precisely and very fast. First, the squared
matrix element is evaluated at $n_c$ different points in the parameter space.
The structure of Eq.~\eqref{eq:morphing} together with some linear algebra is
then used to exactly interpolate to any other parameter point. This process is
described in detail in Ref.~\cite{Brehmer:2018eca}.

\madminer{} implements this morphing technique and leverages it extensively. The
user only has to specify the maximal powers with which each model parameter
contributes to the squared matrix element. \madminer{} then automates the
necessary linear algebra internally.

A practical question is at which $n_c$ benchmark points the matrix elements
should be evaluated originally. This set of parameter points is called the
\emph{morphing basis}. While the physical predictions for a given parameter
point are independent of this basis, the morphing procedure involves matrix
inversions and cancellations between potentially large terms that depend on the
choice of basis.  Some morphing basis choices can thus lead to floating-point
precision issues, while others are numerically more stable. \madminer{} can
automatically pick or complete a morphing basis that avoids or minimizes
numerical precision issues. This optimization consists of randomly drawing a number
of basis configurations over a user-specified parameter region, calculating
morphing weights for each basis, and choosing the
basis that minimizes the sum of squared morphing weights.  

Note that \madminer{} is not restricted to problems that factorize according to
Eq.~\eqref{eq:morphing}. Much of the core functionality is available for almost
any model of new physics. But some features are currently only implemented in
the morphing case, and for others the morphing setup can reduce the
computational cost substantially.

\item[Parton shower]
Parton shower and hadronization can be simulated with
\toolfont{Pythia~8}~\cite{Sjostrand:2007gs}, including matching and merging of
different final-state jet multiplicities. This part of the event evolution
should not directly depend on the new physics parameters of
interest.\footnote{Fundamentally, the presented inference techniques also
support new physics effects that affect \eg the probabilities of shower
splittings, but this is currently not supported in \madminer.} Other shower
simulators can be interfaced with little effort.

\item[Detector simulation]
Out of the box, \madminer{} includes a fast phenomenological detector simulation
with \toolfont{Delphes~3}~\cite{deFavereau:2013fsa}, as well as an alternative
approximate detector simulation through smearing functions based on the
parton-level final state. \madminer{} is designed modularly so that it can be
interfaced to more realistic detector simulations used by the experimental
collaborations such as \toolfont{Geant4}~\cite{Agostinelli:2002hh}. Such an
extension will mostly require careful book-keeping of event weights and
observables.

\item[Observables]
The observed data for each event needs to be parameterized in a fixed-length
vector of observables $x$. These can include both basic characteristics like
energies, transverse momenta, and angular directions of reconstructed particles,
but also higher-level features such as invariant masses or angular correlations
between particles. For \toolfont{Delphes}-level analyses, \madminer{} allows the
definition of these observables as arbitrary functions of the objects in the
\toolfont{Delphes} output file, while for parton-level analyses arbitrary
functions of the smeared parton-level four-momenta are supported. It is possible
to extend \madminer{} with interfaces to any external code that calculates
observables from generated events.

\item[Backgrounds]
Different signal and background processes, with no limitations on the
parton-level final states, can be combined in the same analysis. Background
processes are allowed to depend on the model parameters $\theta$. In the case
of a reducible backgrounds that are not affected by $\theta$, the joint log
likelihood ratio and joint score of all background events are zero, up to an
$x$-independent constant that is related to the dependence of the overall signal
cross section on $\theta$. We will discuss and illustrate this case in
Sec.~\ref{sec:example-illustration}.

While fully data-driven backgrounds are not supported, a data-driven
normalization of MC event samples is possible.

\item[Systematic uncertainties]
All imperfections in the description of the physics process with the simulation
chain are modeled with nuisance parameters $\nu$. For most of the analysis
chain, they play the same role as the physics parameters of interest $\theta$:
their true value is unknown and they affect the likelihood of simulation
outcomes. For the inference techniques presented in the previous section, every
occurence of $\theta$ then has to be replaced with $(\theta, \nu)$: the neural
networks estimate the likelihood $p(x | \theta, \nu)$, the likelihood ratio $r(x
| \theta_0, \nu_0 ; \theta_1, \nu_1)$, or the score $t(x | \theta, \nu)$, where
the latter now has more components corresponding to both the gradient with
respect to $\theta$ and the gradient with respect to $\nu$. At the final limit
setting stage, one then picks a constraint term (or, in a Bayesian setting, a
prior) for the nuisance parameters and profiles (or marginalizes) over them,
following established statistical procedures~\cite{Cowan:2010js,
Brehmer:2016nyr, Edwards:2017mnf, Alsing:2019dvb}.

\madminer{} currently supports nuisance parameters that model systematic
uncertainties from scale and PDF choices. The effect of the nuisance parameters
on an event weight is parameterized as
\be
  \diff \sigma(z_p | \theta, \nu) = \diff \sigma(z_p | \theta, 0) \times
  \exp \left[ \sum_i \left( a(z_p) \, \nu_i + b(z_p) \, \nu_i^2 \right) \right] \,,
  \label{eq:nuisance_model}
\ee
similar to \toolfont{HistFactory}~\cite{Cranmer:2012sba} and
\toolfont{PyHF}~\cite{PYHF}. $\nu_i = 0$ corresponds to the nominal value of the
$i$-th nuisance parameter. For each varied scale, $\nu_i = \pm 1$ correspond to
the scale variations (typically half and twice the nominal scale choice). PDF
uncertainties are described by one nuisance parameter per eigenvector in a
Hessian PDF set, and $\nu_i = 1$ corresponds to the event weight along a unit
step of an eigenvector. The factors $a(z_p)$ and $b(z_p)$ are automatically
calculated for each event based on a reweighting
procedure~\cite{Frederix:2011ss}. The exponential form of
\equref{nuisance_model} ensures non-negative event weights.

\item[Neural network architectures and training]
The heart of \madminer's analysis techniques are neural networks that take event
data $x$ (and, depending on the method, a parameter point $(\theta, \nu)$) as
input and return the likelihood, likelihood ratio, or score. The optimal
architecture of these networks depends on the problem. \madminer{} currently
supports fully connected feed-forward neural networks with a variable number of
layers and hidden units and different activation functions, implemented in
\toolfont{PyTorch}~\cite{paszke2017automatic}. The loss functions are mostly
fixed by the inference methods given in Tbl.~\ref{tbl:inference_techniques}. The
\scandal, \rascal, \alices, and \cascal techniques have a free hyperparameter
$\alpha$ that weights the joint score term in the loss function relative to
another term. These loss functions are minimized by stochastic gradient descent
with or without momentum~\cite{Qian:1999:MTG:307343.307376}, the \toolfont{Adam}
optimizer~\cite{2014arXiv1412.6980K}, or the \toolfont{AMSGrad}
optimizer~\cite{j.2018on}; other options include batching, learning rate decay,
and early stopping.

\item[Uncertainty estimation]
An individual neural estimator merely provides a point estimate for the
likelihood, likelihood ratio, or score. By training an ensemble of different
estimators with different random seeds, we can use the ensemble variance as a
diagnostic tool to check whether the global minimum of the loss functional has
been found~\cite{2016arXiv161201474L}. Taking this idea one step further, we can
train each network on resampled data. With this nonparametric bootstrap method,
the ensemble variance represents the uncertainty in the neural network output
from finite training sample size. While this approach may serve as a useful
indicator of the epistemic uncertainty of the network predictions (\ie the
uncertainty on the parameters of the neural network), there is no guarantee that
it covers all relevant sources of bias and variance.
%
\end{description}

\subsection{Recommendations for getting started}
\label{sec:recommendations}

The large number of different inference methods, analysis aspects, and
hyperparameters outlined above and described in detail in a total of six
publications~\cite{Cranmer:2015bka, 2017arXiv170507057P, Brehmer:2018hga,
Brehmer:2018kdj, Brehmer:2018eca, Stoye:2018ovl} might seem a little
overwhelming. That is why we here provide a few suggestions for new users of
\madminer, largely based on the comprehensive comparison in
Refs.~\cite{Brehmer:2018eca, Stoye:2018ovl}. Rather than being a
one-size-fits-all solution, this should be seen as a starting point for the
exploration of the space of possible analysis methods.

The main question is whether one of the methods should be used that reconstruct
the entire likelihood or likelihood ratio function, as described in
Sec.~\ref{sec:likelihood_estimators}, or whether the analysis merely aims to
find (locally) optimal observables, as described in
Sec.~\ref{sec:score_estimators}. The former approach is potentially more
powerful: given enough data, expressive enough networks, and a training of the
neural network that reaches the global minimum of the loss function, it will
lead to the best possible limits. But it is also more ambitious, may require
more training data and hyperparameter experiments, and represents a bigger
change to a typical data analysis pipeline.

The latter strategy, on the other hand, is simpler, scales better to
high-dimensional parameter spaces, and requires less training samples. Since it
essentially defines a new set of observables, it requires only minimal
modifications to existing analysis pipelines. The Fisher information formulation
makes it very easy to summarize the sensitivity of a measurement. The catch is
that this approach is only optimal as long as the dominant signatures enter at
linear order in the model parameters, and otherwise loses statistical power and
may lead to worse limits.

If this last condition is satisfied\,---\,if the dominant new physics effects
are expected at linear order in the parameters\,---\,we consider the \sally
strategy an ideal starting point. A typical example for this is a precision
measurement of effective operators. On the other hand, if nonlinear
contributions from the model parameters dominate, we instead suggest using the
\alices technique. Its hyperparameter $\alpha$ should initially be chosen such
that the two terms in the loss function contribute approximately equally to the
training, but it is worth scanning this parameter over a few orders of
magnitude.

\section{Using MadMiner}
\label{sec:implementation}

We will now describe the implementation of these techniques in the new Python
package \madminer{}.

\madminer{} is open source and its code is available at
Ref.~\cite{MadMiner_repo}. That repository also contains interactive tutorials
with step-by-step comments. A detailed documentation of the API is available
online at Ref.~\cite{MadMiner_docs}. We also provide a \toolfont{Docker}
container with a working environment of all required tools at
Ref.~\cite{MadMiner_docker}, and reusable workflows based on
\toolfont{Reana}~\cite{Simko:2652340} at Ref.~\cite{MadMiner_reana}.

To get started, the minimal requirements are working installations of
\toolfont{MadGraph5\_aMC} and \madminer{}. The latter can be installed with a
simple \pyth{pip install madminer}. Shower and detector simulations in addition
require installations of \toolfont{Pythia~8}, the automatic MadGraph-Pythia
interface, and \toolfont{Delphes~3}. To model PDF uncertainties,
\toolfont{LHAPDF} has to be installed, including its Python interface. All these
additional dependencies can easily be installed from the
\toolfont{MadGraph5\_aMC} command line interface. Detailed instructions for the
installation can be found at Ref.~\cite{MadMiner_docs}.

\begin{figure*}
\centering
\includegraphics[width=0.885\textwidth]{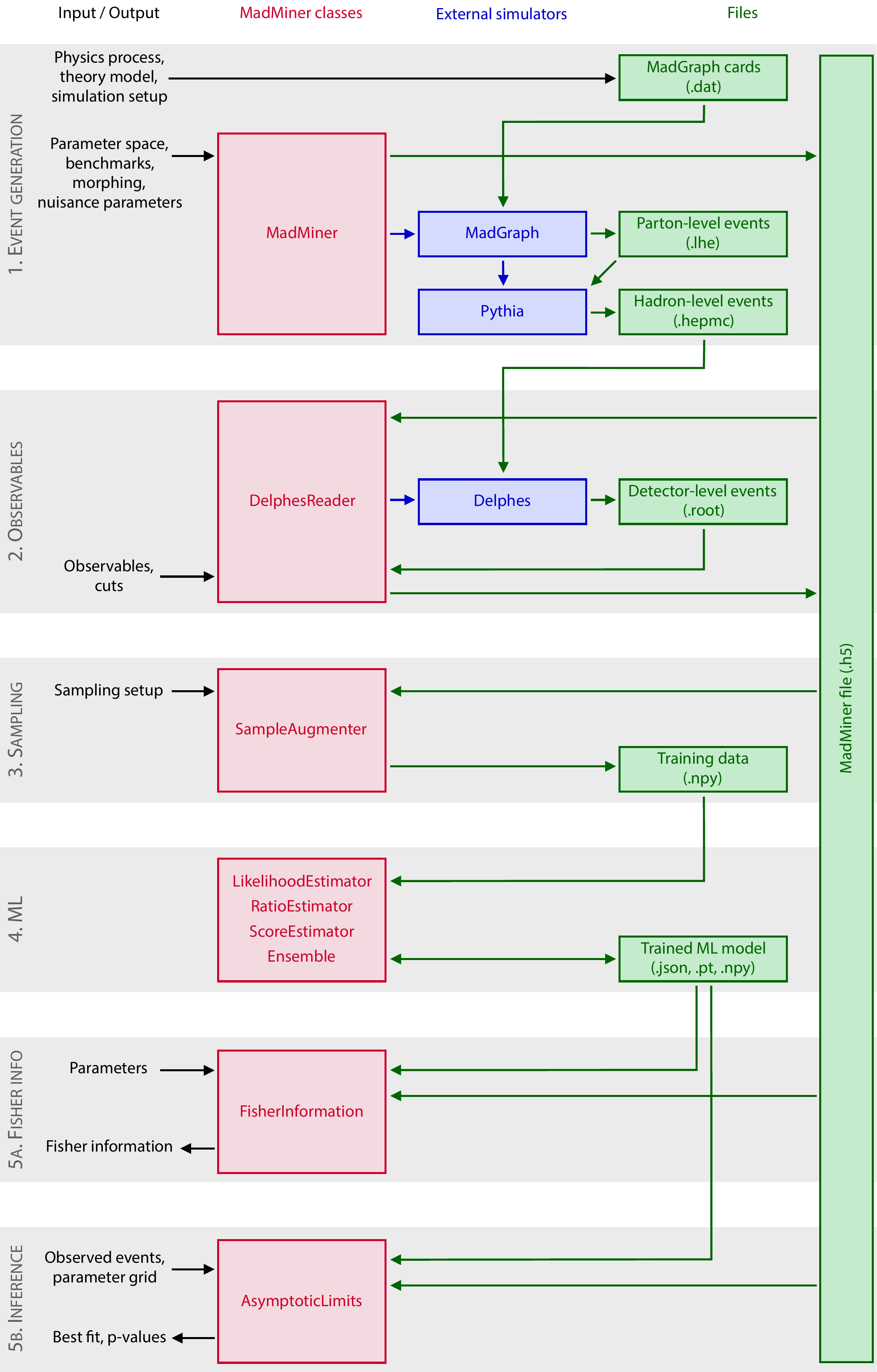}%
\caption{Example workflow, with classes in red, external simulations in blue,
and files in green.}
\label{fig:workflow}
\end{figure*}

In the following we will go through the typical steps of a \madminer{} analysis
that uses the inference techniques discussed in the last section.
Figure~\ref{fig:workflow} visualizes the workflow of such an analysis, and we
will generally follow this figure.

\subsection{Analysis specification and event generation}

The first phase of a \madminer analysis consists of specifying the problem and
generating events. First, the necessary files (``cards'') that define the
analyzed process and theoretical model should be collected. This includes the
UFO model files as well as the run card, the parameter card, the
\toolfont{Pythia} card, and the \toolfont{Delphes} card, all in the standard
format used by \toolfont{MadGraph5\_aMC}.

The measurement problem is specified with an instance of the \pyth{MadMiner}
class. The parameter space is defined by repeatedly calling its
\pyth{add_parameter()} function. Each model parameter is specified by its LHA
block and LHA ID in the UFO model.

Next, the user chooses \emph{benchmarks}: parameter points at which
the event are evaluated. Benchmarks can be specified manually with
\pyth{add_benchmark()}. Additionally, a morphing technique based on
Eq.~\ref{eq:morphing} can be activated by calling
\pyth{set_morphing()}. If less benchmarks have been manually specified than
required for a morphing basis,
more benchmarks will be chosen automatically,
minimizing the expected size of morphing weights $|w_c(x)|$.

Systematic uncertainties (from PDF and scale variations) can be specified with a
call to \pyth{set_systematics()}. Once the parameter space, benchmarks,
morphing, and systematic uncertainties are set up, \pyth{save()} saves these
settings in the \emph{MadMiner file}, which is based on the HDF5
standard~\cite{hdf5}.

Finally, events can be generated by calling \pyth{run()} or
\pyth{run_multiple()} (the difference is that the former starts one
event generation run, while the latter generates multiple sets with different
run cards or sampled from different benchmarks). \madminer{} will set up
\toolfont{MadGraph}'s reweighting feature to evaluate the event weights for all
events at all benchmarks, which are stored in the LHE event files together with
the parton-level information. \toolfont{Pythia~8} will automatically be called
to shower and hadronize the partons, the results are stored in a standard HepMC
event file~\cite{Dobbs:2001ck}.

\subsection{Detector effects and observables}

In the second phase, all relevant information has to be extracted from the event
samples, including observables as well as event weights for the different
benchmarks. There are currently two implementations for this step: the
\pyth{LHEReader} class realizes a simple parton-level analysis, in which the
effects of shower and detector are approximated with transfer functions, while the
\pyth{DelpesReader} class implements an detector-level analysis in which the
shower is modeled with \toolfont{Pythia~8} and the detector with
\toolfont{Delphes~3}. The API of both classes is very similar, here we focus on
the \pyth{DelpesReader} option.

After creating a \pyth{DelphesReader} instance and pointing it to the
\madminer{} file, the user has to list the HepMC event samples that should be
analyzed by calling the function \pyth{add_sample()}. The detector simulation
with \toolfont{Delphes} can either be run externally or through \madminer{} by
calling \pyth{run_delphes()}.

In a next step, the user defines a set of observables that will be calculated
for each event. These can be provided either as Python functions with
\pyth{add_observable_from_function()} or as parse-able strings with
\pyth{add_observable()}. In both cases, reconstructed objects are accessible as
\pyth{MadMinerParticle} objects, which inherits all functions of
\toolfont{scikit-hep}'s \pyth{LorentzVector}
class~\cite{eduardo_rodrigues_2019_3234683}. This makes observable definitions
very easy: for instance, the transverse momentum of the hardest lepton can
simply be defined as \pyth{add_observable("lepton_pt", "l[0].pt")}, while the
azimuthal angle between the two hardest jets can be defined as
\pyth{add_observable("delta_phi", "j[0].deltaphi(j[1])")}. Cuts can be added
similarly with \pyth{add_cuts()}.

Once all samples are added, \toolfont{Delphes} has been called, and all
observables and cuts are defined, a call to \pyth{analyse_delphes_samples()}
parses the observables for the simulated events, applies the cuts, and extracts
the relevant event weights. With \pyth{save()} this data is stored in the
\madminer{} file.

\subsection{Sample unweighting and augmentation}

If multiple different samples were created, for instance for different processes
or phase-space regions, they should now be combined into a single \madminer{}
file and shuffled by calling the \pyth{combine_and_shuffle()} function.

In the third step of the analysis workflow, the event information in the
\madminer{} file is processed into training data for the different algorithms
described in the previous section. This consists of two aspects: first, the
event data needs to be reweighted to the parameter points $\theta$ (and\,/\,or
$\theta_0$, $\theta_1$, $\thetaref$) that make up the training data and then
unweighted. Second, the joint likelihood ratio $r(x,z)$ and the joint score
$t(x,z)$ need to be calculated for each unweighted event.

This is implemented in the \pyth{SampleAugmenter} class. It provides a set of
six high-level functions that generate and augment the data for the different
types of inference techniques. For instance, \pyth{sample_train_local()}
generates training samples for score estimators (the \sally and \sallino
techniques), while \pyth{sample_train_ratio()} prepares training data for
likelihood ratio estimators. The output of all these functions are a set of
plain \toolfont{NumPy}~\cite{numpy} arrays. The rows of these two-dimensional
arrays are the events; the columns correspond to the
observables that characterize the event data (in the order in which the
observables were defined in the \pyth{DelphesReader} or \pyth{LHEReader}
classes), the parameter points according which they are sampled, and
the components of the joint score, respectively.

\subsection{Machine learning}

It is finally time to train neural networks to estimate the likelihood,
likelihood ratio, or score, as discussed in Sec.~\ref{sec:inference}. This is
implemented in the classes \pyth{LikelihoodEstimator},
\pyth{ParameterizedRatioEstimator}, \pyth{DoubleParameterizedRatioEstimator},
and \pyth{ScoreEstimator}. This training is independent of the external
Monte-Carlo simulations and even the \madminer{} file, which makes it easy to
run it on an external system with GPU support.

During initialization of any of these classes, the network architecture is
chosen. Currently, \madminer{} supports fully connected feed-forward networks
with variable number of layers, hidden units, and activation functions,
implemented in \toolfont{PyTorch}~\cite{paszke2017automatic}. A call to
\pyth{train()} starts the training; keywords specify which loss function to use,
the location of the training data generated in the previous step, the optimizer,
the learning rate schedule, the batch size, and whether early stopping is used.

After training, \pyth{save()} saves the neural network to files. The estimators
are evaluated for arbitrary parameter points and observables with
\pyth{evaluate_log_likelihood()}, \pyth{evaluate_log_likelihood_ratio()}, or
\pyth{evaluate_score()}.  For many users, the estimates returned by these
functions will be the final output of \madminer{}, and the statistical analysis
will be performed externally.

We also provide the \pyth{Ensemble} class, a convenient wrapper that allows to
train an ensemble of multiple neural networks. The different instances can have
identical or different architectures and the training can be performed on the
same or resampled training data. Such an ensemble is useful for consistency
checks and uncertainty estimation as discussed in
Sec.~\ref{sec:practical_analysis_aspects}.

\subsection{Inference}

\madminer{} provides a barebones framework for the statistical analysis:
the \pyth{AsymptoticLimits} class. After initalizing it with the \madminer{}
file, the two high-level functions \pyth{expected_limits()} and
\pyth{observed_limits()} calculate expected and observed $p$-values over a grid
in parameter space.  \pyth{expected_limits()} takes as input the parameter point
that is assumed to be true and internally generates a so-called ``Asimov'' data
set~\cite{Cowan:2010js}, a large simulated set of events.
\pyth{observed_limits()} on the other hand is directly based on a list of
events, which the user can take from simulations or actual measured data.

Both methods can estimate the kinematic likelihood either through histograms of
kinematic variables, through histograms of the estimated score from a trained
\pyth{ScoreEstimator} instance, or through a trained likelihood (ratio)
estimator. $p$-values are calculated with a likelihood ratio test, using
the asymptotic distribution of the likelihood ratio as described in Wilks'
theorem~\cite{Wilks:1938dza, Wald, Cowan:2010js}.

The \pyth{AsymptoticLimits} currently does not support systematic uncertainties.
We are planning to interface \madminer{} with existing software packages that
implement profile likelihood ratio tests.

\subsection{Fisher information}

As discussed in Sec.~\ref{sec:score_estimators}, a convenient and powerful
summary of the sensitivity of a measurement is the Fisher information matrix.
Its calculation is implemented in the \pyth{FisherInformation} class. Most
importantly, \pyth{full_information()} calculates the Fisher information based
on a \pyth{ScoreEstimator} instance as given in Eq.~\eqref{eq:fisher_info_lhc}.
Several other functions allow to calculate the Fisher information in the cross
section only (\ie the first term of Eq.~\eqref{eq:fisher_info_lhc}),
the Fisher information in the histogram of one or two kinematic
variables, and finally the truth-level Fisher information, which treats all
properties of the parton-level particles as observable. Finally, the function
\pyth{histogram_of_information()} allows the user to calculate the distribution
of the Fisher information over phase space, as introduced in
Ref.~\cite{Brehmer:2016nyr}.

In the presence of systematic uncertainties and in a frequentist setup, nuisance
parameters can either be neglected (``projected out'') or conservatively taken
into account (``profiled out''). These operations are implemented in the
functions \pyth{project_information()} and \pyth{profile_information()}.

\section{Physics example}
\label{sec:example}

We demonstrate the use of \madminer in the measurement of dimension-six
operators in $tth$ production at the high-luminosity run of the LHC. We choose
to analyze fully leptonic top decays and a Higgs decay into two photons,
\be
  p p \to t \bar{t} \, h \to (b \ell^+) \, (\bar{b} \ell^-) \, (\gamma \gamma) \, \met
\ee
with $\ell = e, \mu$. While this particular signature is not expected to be the
most sensitive channel, for example when compared to either semi-leptonic $tth$
production or Higgs production in gluon fusion, it provides a high-dimensional
final state with a non-trivial missing energy signature, illustrating the
features and challenges that \madminer{} can address.

We consider three different scenarios:
\begin{description}
  \item[Illustration]
  We first illustrate the mechanism behind the inference techniques in
  \madminer{} in a one-dimensional version of the problem, restricting the
  analysis to one parameter and one observable.

  \item[Validation]
  \madminer{} is then validated in a parton-level toy scenario. By not letting
  the $W$ bosons decay and ignoring the effect of shower and detector on
  observables, we can calculate the true likelihood function and compare the
  output of the neural networks to a ground truth.

  \item[Physics analysis]
  Finally, we perform a realistic phenomenological analysis, including the
  effects of parton shower and detector and considering a three-dimensional
  parameter space and high-dimensional event data.

\end{description}

All three analyses are performed with \toolfont{MadMiner~v0.4} following the
workflow outlined in the previous section. Events are generated with
\toolfont{MadGraph5\_aMC@NLO} at leading order for $\sqrt{s} = 14~\tev$ using
the \toolfont{PDF4LHC15\_nlo} parton distribution
function~\cite{Butterworth:2015oua}. We normalize the rates to the NLO
predictions~\cite{deFlorian:2016spz} with a phase-space-independent $k$-factor.
We consider the Standard Model Lagrangian supplemented with dimension-six
operators in the SILH basis~\cite{Giudice:2007fh}, as implemented in the
\toolfont{HEL} \toolfont{FeynRules} model~\cite{Alloul:2013naa}.

\begin{table}
\begin{tabular*}{0.95\textwidth}{@{\extracolsep{\fill}} l ccc}
  \toprule
  & Illustration & Validation & Physics analysis \\
  \midrule
  Operators & $\ope{G}$ & $\ope{u}$, $\ope{G}$, $\ope{uG}$ & $\ope{u}$, $\ope{G}$, $\ope{uG}$ \\
  Initial states & $pp$ & $gg$ & $pp$ \\
  Final state & $(b \ell^+) \, (\bar{b} \ell^-) \, (\gamma\gamma) \, \met$
  &$(bW^+) \, (\bar{b}W^-) \, (\gamma \gamma)$
  & $(b \ell^+) \, (\bar{b} \ell^-) \, (\gamma\gamma) \, \met$ \\
  Background & $\checkmark$ & -- & $\checkmark$ \\
  Shower simulation & \toolfont{Pythia} & -- & \toolfont{Pythia} \\
  Detector simulation & \toolfont{Delphes} & -- & \toolfont{Delphes} \\
  Observables & 1: $p_{T,\gamma\gamma}$ & 80 & 48 \\
  Systematic uncertainties & -- & -- & PDF, scale \\
  \bottomrule
\end{tabular*}
\caption{The three scenarios in which we analyze the $tth$ process.
}
\label{tbl:analyses}
\end{table}

Otherwise, the simulation setup is different for each of the three scenarios. We
summarize the main settings in Tbl.~\ref{tbl:analyses} and discuss them in each
of the following sections.

\subsection{Illustration of analysis techniques}
\label{sec:example-illustration}

Our first analysis aims to illustrate how \madminer{} calculates the likelihood
function in a simplified one-dimensional version of the problem. For this we
restrict ourselves to a single dimension-six operator,
\be
  \mathcal{L} = \mathcal{L}_{SM} + c_G \, \ope{G}
\ee
with
\be
  \ope{G} = \frac{g_s^2}{m_W^2} (H^\dagger H) G_{\mu\nu}^a G_a^{\mu\nu} \,.
\ee
This operator induces an additional contribution to the effective Higgs-gluon
coupling, $g_{ggh} \to g_{ggh} (1+  192 \pi^2/g^2  \times  c_G)$, and therefore
affects the kinematic distributions~\citep{Maltoni:2016yxb}.

We define the theory parameter as $\theta = 100\,c_G$, which is dimensionless
and of order unity over the parameter range of interest. $\theta = 0$ then
corresponds to the SM, any deviation from zero to a new physics effect. The
squared matrix element consists of an SM contribution, an interference term
linear in $\theta$, and a squared dimension-six amplitude proportional to
$\theta^2$, and we can use a morphing technique to interpolate event weights and
cross sections from three benchmarks (or morphing basis points) to any point in
parameter space.

In this illustration setup we also restrict the analysis to a single observable
$x=p_{T,\gamma\gamma}$, the transverse momentum of the di-photon system. All
other observables are treated as if they were unobservable. Together with
physically unobservable degrees of freedom (such as neutrino energies) as well
as random variables in the simulation of the shower, hadronization, and
detector, they form the set of latent variables $z$. This setup is similar to a
histogram-based analysis of $c_G$ using only the $p_{T,\gamma\gamma}$ histogram.

We generate events with \toolfont{MadGraph5\_aMC@NLO} as described above. They
are then showered and hadronized through \toolfont{Pythia~8}. The detector
response is simulated with \toolfont{Delphes~3} using the HL-LHC card suggested
by the HL/HE-LHC working group~\cite{Cepeda:2019klc}.

\subsubsection{Signal only}

\begin{figure*}
  \centering
  \includegraphics[width=0.49\textwidth]{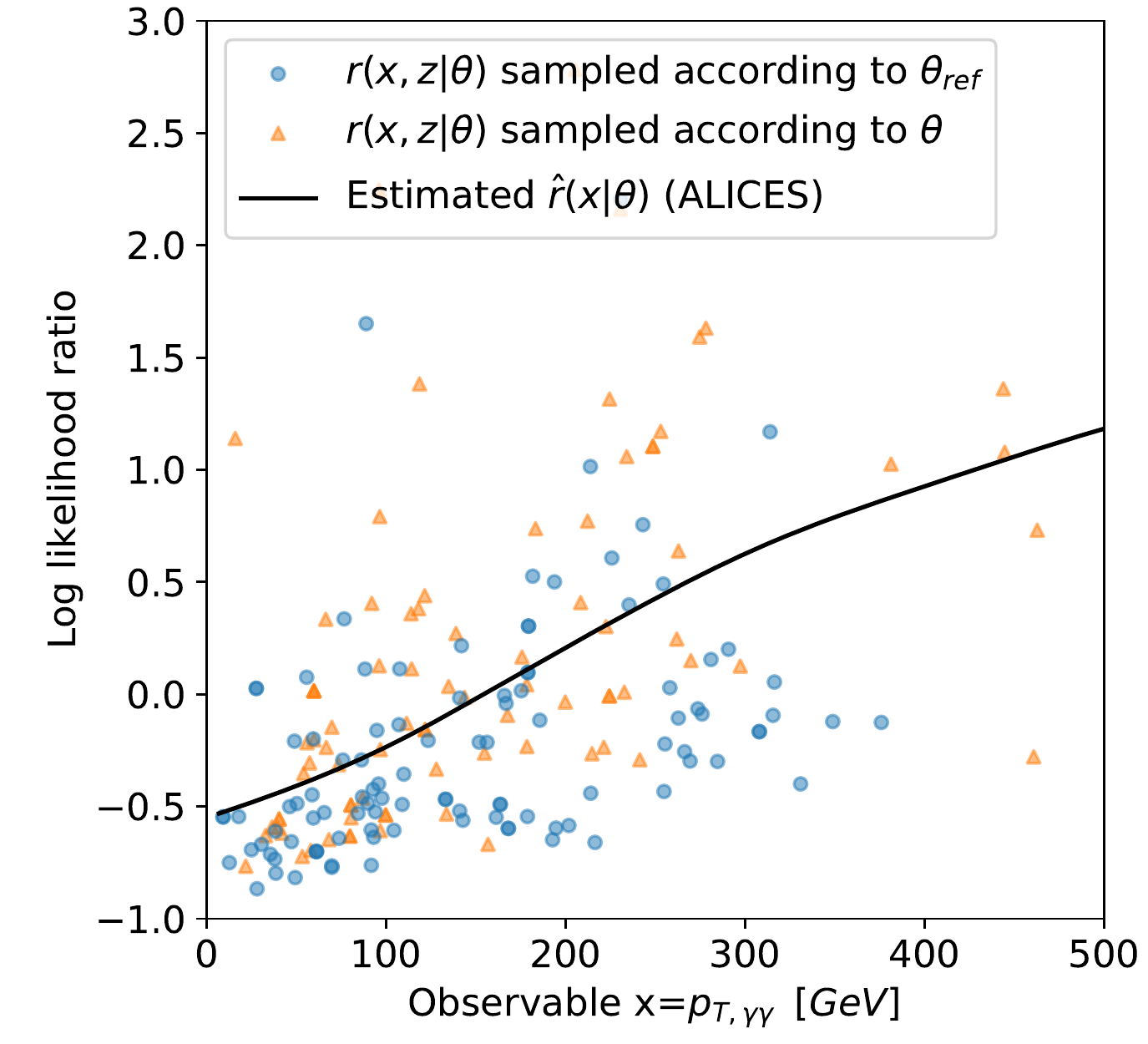}%
  \includegraphics[width=0.49\textwidth]{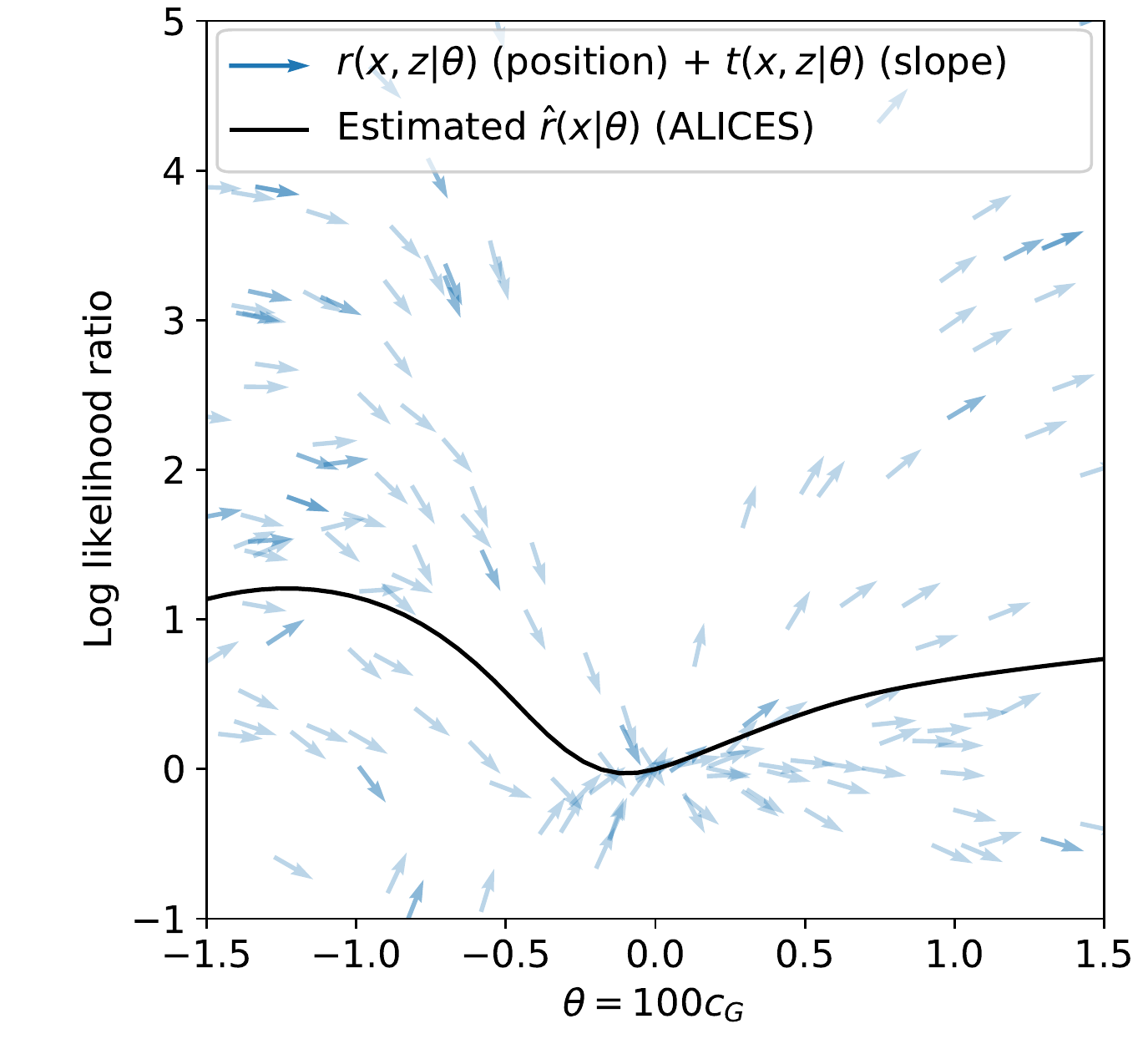}%
  \caption{Illustration of the analysis techniques in a one-dimensional problem.
  \textbf{Left}:
  Joint log likelihood ratio as a function of the observable
  $p_{T, \gamma\gamma}$ for $tth$ signal events sampled according to the SM (blue
  dots) and a BSM theory with $\theta=100\,c_G=1$ (orange triangles). The solid line
  shows the estimated log likelihood ratio from an \alices model trained only on
  $p_{T\gamma\gamma}$ as input observable.
  \textbf{Right}:  Joint log likelihood ratio (arrow position) and joint score
  (arrow slope) as a function of the model parameter $\theta=100\,c_{G}$, for
  $tth$ signal events in the range $p_{T,\gamma\gamma}=(300\pm 2.5)$~GeV. The solid
  line shows the estimated  log likelihood ratio from an \alices model trained
  only on $p_{T,\gamma\gamma}$ as input observable and evaluated at
  $p_{T,\gamma\gamma}=300$ GeV.}
  \label{fig:illustration}
  \end{figure*}

In the sampling and data augmentation step (the third box in \figref{workflow}),
\madminer{} creates training samples where each simulated event is characterized
by values of the observable $x=p_{T,\gamma\gamma}$ and the (unobservable) latent
variables $z$. Additionally, for each event \madminer calculates the joint
likelihood ratio $r(x,z|\theta)$ between the parameter point $\theta$ and a
reference point $\thetaref$, which we take to be the SM. It also calculates the
joint score $t(x,z|\theta)$ evaluated at the parameter point $\theta$.
This is illustrated in \figref{illustration}. The blue dots and orange triangles in the
left panel show the joint log likelihood
ratio $\log r(x,z|\theta)$ with their dependence on the observable $x =
p_{T,\gamma\gamma}$. The blue dots show $tth$ events sampled according to the SM
(with $\thetaref = 0$), while the orange triangles are sampled from a BSM hypothesis
with $\theta=1$ (or $c_G=0.01$). We can see that there are more high-$p_{T}$
events for the BSM model than for the SM, and hence the joint likelihood ratio
is higher. The large vertical scatter in the joint likelihood ration is caused
by the presence of the latent variables $z$, which affect the joint likelihood
ratio, but are unobservable. In the right panel of the same figure, the arrows
show the joint log likelihood ratio $\log r(x,z|\theta)$ (arrow position) and
the joint score $t(x,z|\theta)$ (arrow slope) with their dependence on the
theory parameter $\theta$. Here the observable is constrained to the range
$p_{T,\gamma\gamma}=(300 \pm 2.5)~\gev$ to suppress the observable dependence.

Estimating the likelihood ratio with the methods described in
Sec.~\ref{sec:likelihood_estimators} (and in more detail in
Ref.~\cite{Brehmer:2018eca}) essentially means fitting a function
$\hat{r}(x|\theta)$ to the joint likelihood ratio $r(x,z|\theta)$ by numerically
minimizing a suitable loss functions. In this process, the unobservable latent
variables $z$ are effectively integrated out. This is the gist of the machine
learning step of the \madminer{} workflow (box four in \figref{workflow}). The
result of this step, the estimated  log likelihood ratio $\hat{r}(x|\theta)$
based on the \alices method, is shown in the solid black lines in
\figref{illustration}: the left panel illustrates the $x$ dependence for fixed
$\theta=1$, the right panel the $\theta$ dependence for fixed $x$. While it is
possible to estimate the likelihood ratio only using the joint likelihood ratio
as input, the gradient information that is the joint score provides additional
guidance, which often allows for the fit to converge with less data.

\subsubsection{Adding backgrounds}

\begin{figure*}
\centering
\includegraphics[width=0.49\textwidth]{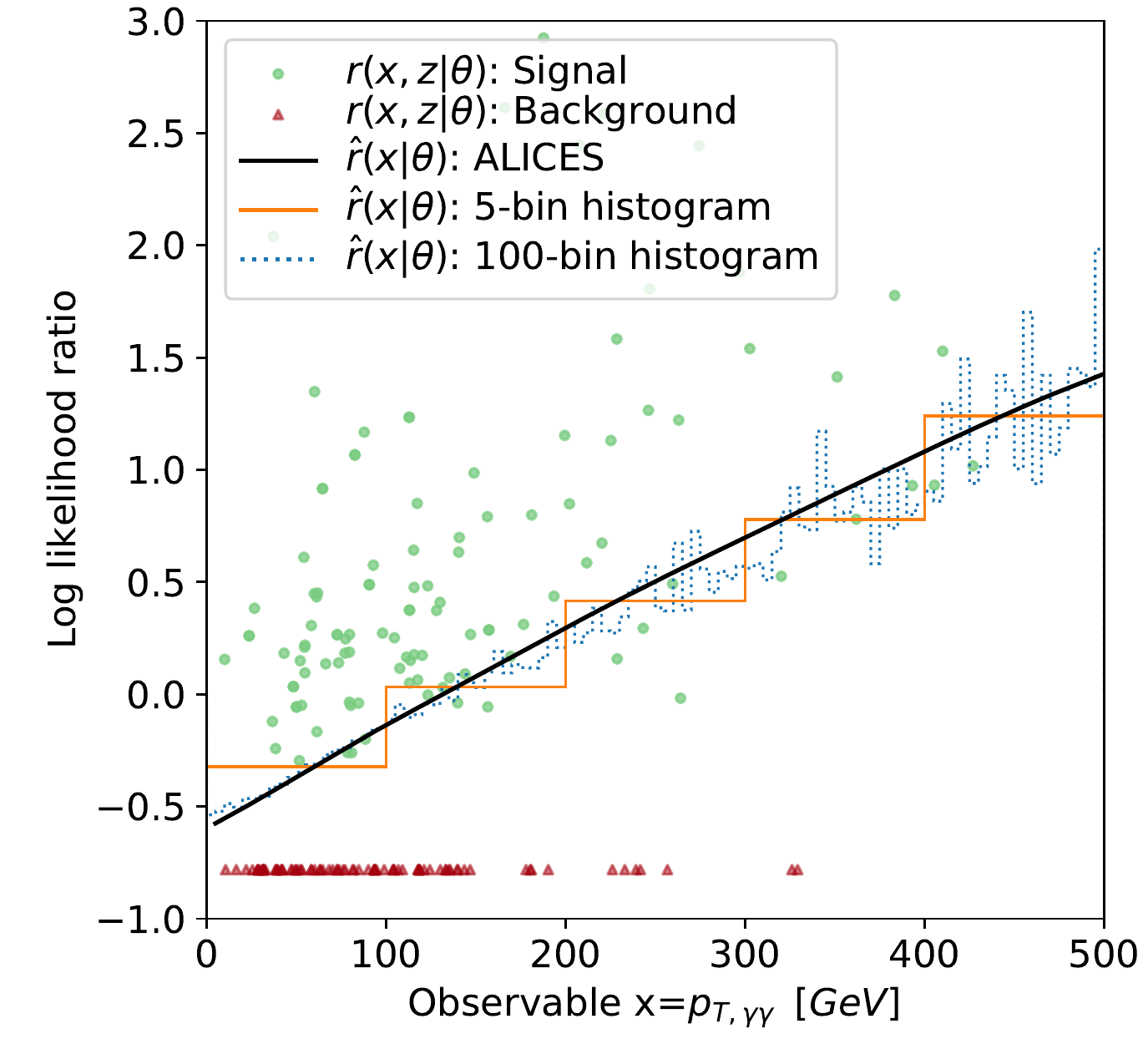}%
\includegraphics[width=0.49\textwidth]{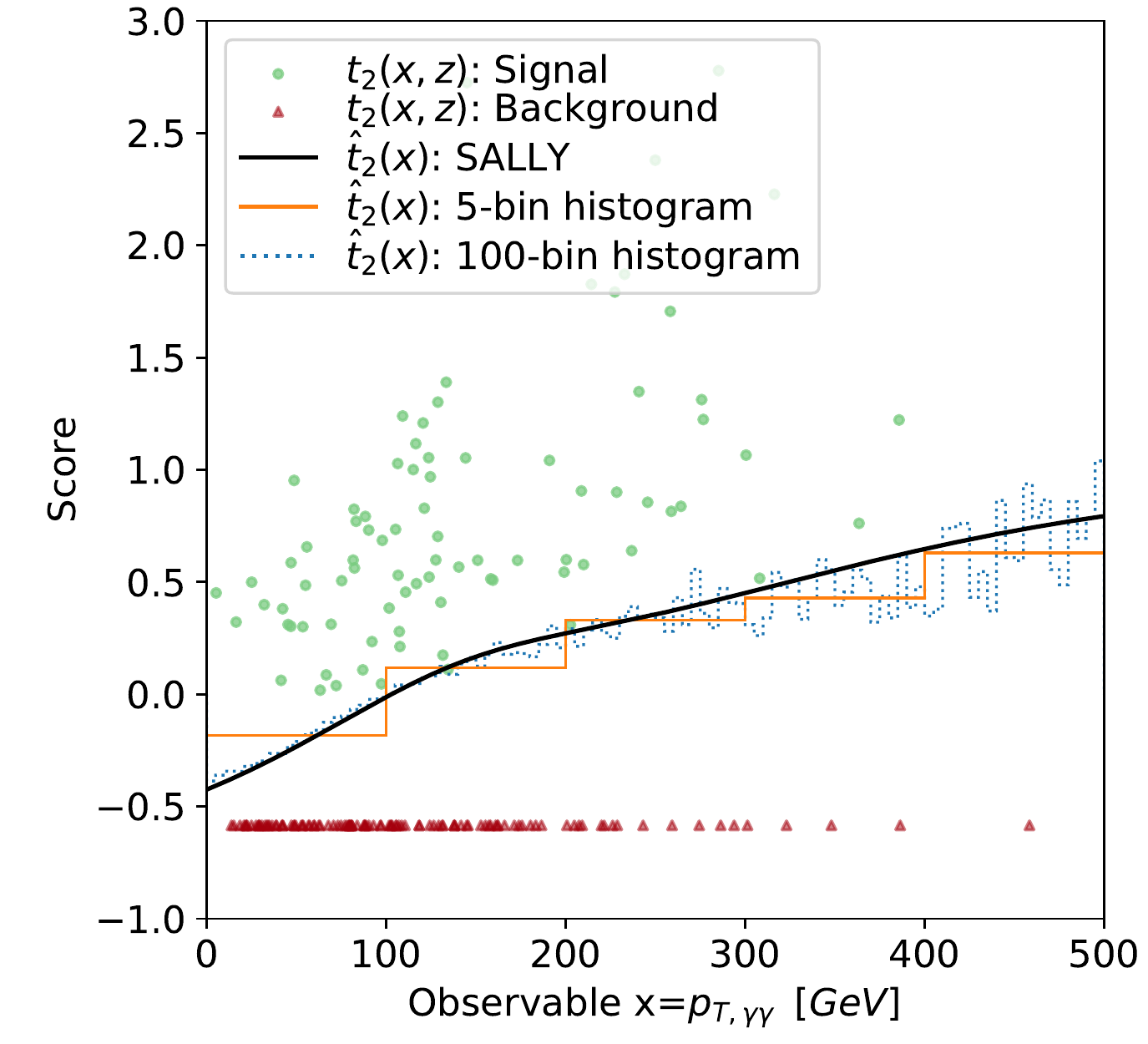}%
\caption{Illustration of the analysis techniques in a one-dimensional problem.
\textbf{Left}: Joint log likelihood ratio as a function of the observable
$p_{T\gamma\gamma}$ for $tth$ signal events (green dots) and $tt\gamma\gamma$
background events (red triangles) sampled according to the SM and a BSM theory with
$\theta=100~c_G=1$. The background events cluster at a constant value of
$-0.78$, as explained in the text. The lines show the estimated log likelihood
ratio based on the \alices method trained only on $p_{T,\gamma\gamma}$ (black
solid) and a $p_{T,\gamma\gamma}$ histogram with 5 (orange solid) and 100 (blue
dashed) bins, respectively.
\textbf{Right}: Joint score evaluated at the SM for for $tth$ signal (green
dots) and $tt\gamma\gamma$ background events (red triangles). The background events
cluster at a constant value of $-0.58$, as explained in the text. The lines show
the estimated score obtained using a \sally method trained only on
$p_{T\gamma\gamma}$ (black solid) and a $p_{T,\gamma\gamma}$ histogram with 5
(orange solid) and 100 (blue dashed) bins, respectively.}
\label{fig:illustration2}
\end{figure*}

So far we have only considered the $tth$ signal process. How does this picture
change when we include backgrounds? We answer this question in the left panel of
\figref{illustration2}, where in addition to the signal we now include the
dominant background, continuum $t\bar{t} \, \gamma\gamma$ production with
leptonically decaying tops.

As before the circles show the joint log likelihood ratio $\log r(x,z|\theta)$
and the line denotes the estimated log likelihood ratio function $\log
\hat{r}(x|\theta)$. Since signal and background populate different phase-space
regions, the interference between them is negligible and we could consistently
simulate them separately from each other. This means that every simulated event
is labeled either as a signal or a background event, which plays the role of a
discrete variable in the set of latent variables $z$. The background event
weights are unaffected by the EFT operator $\ope{G}$, so the joint likelihood
ratio for these events is independent of $x$ and $z$:
\be
  r(x,z|\theta) \bigg|_{\text{background}}
  = \frac{p(x,z|\theta)}{p(x,z|\thetaref) }
  = \frac{\diff \sigma(z_p|\theta)}{\diff \sigma(z_p|\thetaref)} \,
  \frac{\sigma(\thetaref)}{\sigma(\theta)}
  = \frac{ \sigma(\thetaref) }{ \sigma(\theta) } \,,
\ee
which in our case turns out to be
\be
  \log r(x,z|\theta) \bigg|_{\text{background}} = -0.78 =: \log r^* \,.
\ee
This is clearly visible in the left panel of \figref{illustration2}, where the
$t\bar{t} \, \gamma\gamma$ events show up as a horizontal line at this value.
While the presence of backgrounds does not affect the fundamental validity of
the inference technique, it increases the variance of the joint likelihood ratio
around the true likelihood ratio so that more training events are required
before the neural network converges on the true likelihood ratio function.

In this simple example with one-dimensional observations $x$, we can validate
the \alices predictions with histograms. The histogram approximation for the
likelihood ratio is $\hat{r}(x|\theta) = [\sigma_{\text{bin}}(\theta) /
\sigma(\theta)] / [\sigma_{\text{bin}}(\thetaref) / \sigma(\thetaref)]$, where
$\sigma_{\text{bin}}(\theta)$ is the cross section in the bin corresponding to
$x$. In the left panel of \figref{illustration2}, the log likelihood ratio based
on a histogram with 5~(100) equally sized bins is shown as solid orange (dashed
blue) line. It generally agrees excellently with the \alices prediction. The two
histogram lines show the trade-off in the number of bins: while too few bins
lead to large binning effects, a large number of bins can lead to large
fluctuations due to limited Monte-Carlo statistics. In contrast, the \madminer{}
techniques based on neural networks learn the correct continuum limit equivalent
to an infinite number of histogram bins, without suffering from large
fluctuations.

In Sec.~\ref{sec:score_estimators} we described an alternative approach in which
\madminer{} calculates the score, a vector of summary statistics that are
statistically optimal close to a reference parameter point such as the SM. We
illustrate this \sally technique in the right panel of \figref{illustration2}.
The green circles show the joint score $t(x,z)$ at the SM reference point,
corresponding to the change of the log likelihood when infinitesimally
increasing the sole theory parameter $\theta = 100\,c_G$. In analogy to the log
likelihood ratio, the red points clustering at a horizontal line
\be
  t(x,z) \bigg|_{\text{background}} = -0.58 =: t^*
\ee
correspond to the $t\bar{t}\,\gamma\gamma$ background events. Estimating the
score function conceptually corresponds to fitting a function $\hat{t}(x)$ to
the joint score data $t(x,z)$ by numerically minimizing an appropriate loss
function. The resulting score estimator $\hat{t}(x)$ is shown as a solid black
line. Again, in this one-dimensional case we can compare the result to the score
estimated through a histogram, which is shown in a solid red (dashed green) line
for a histogram with 5 (100) bins. We find excellent agreement between the
\sally prediction and the histogram results.

\subsection{Validation at parton level}
\label{sec:example-validation}

Next, we validate \madminer{} in a setup in which we can calculate a ground
truth for the output of the algorithms.  This is not trivial because the ground
truth\,---\,the true likelihood, likelihood ratio, or score\,---\,is intractable
in realistic situations. In the last section we showed how we can use histograms
to check the algorithms, but only when limiting the analysis to one or two
observables. We now turn to another approximation in which we can access the
true likelihood ratio and score, even though both observables and model
parameters are high-dimensional: Following Ref.~\cite{Brehmer:2018kdj,
Brehmer:2018eca}, we consider a truth-level scenario in which all latent
variables are also observable, $x=z$. In this case the likelihood ratio $r(x)$
is equal to the joint likelihood ratio $r(x,z)$ and the score $t(x)$ is equal to
the  joint score $t(x,z)$. We can thus compare the predictions of a neural
network trained to estimate either of these quantities to a ground truth.

For this validation we choose the parton-level process
\be
  gg \to t\bar{t}h \to (bW^+) \, (\bar{b}W^-) \, (\gamma\gamma) \,.
\ee
We do not let the $W$ bosons decay and assume that the four-momenta and flavors
of all initial-state and final-state particles can be measured, \ie we do not
simulate the effect of parton shower and detector response. These truth-level
approximations are not necessary for the inference techniques in \madminer{},
but they allow us to calculate a ground truth for the likelihood ratio and
score, which is not possible for any realistic treatment of neutrinos or
modeling of parton shower and detector response.

Following Ref.~\cite{Maltoni:2016yxb}, we consider three dimension-six operators
affecting the top and Higgs couplings in $tth$ production:
\be
  \mathcal{L} = \mathcal{L}_{SM} + c_u \, \ope{u} + c_G \, \ope{G} + c_{uG} \, \ope{uG} \,,
  \label{eq:lagrangian}
\ee
where the operators are defined as
\be
 \ope{u} = -\frac{1}{v^2} (H^\dagger H) (H^\dagger  \bar{Q}_L) u_R \,, \
 \ope{G} = \frac{g_s^2}{m_W^2} (H^\dagger H) G_{\mu\nu}^a G_a^{\mu\nu}  \,, \
 \ope{uG} = -\frac{4g_s}{m_W^2} y_u  (H^\dagger  \bar{Q}_L)  \gamma^{\mu\nu} T_a u_R G^a_{\mu\nu} \,.
 \label{eq:operators}
\ee

The $\ope{u}$ operator effectively rescales the top Yukawa coupling as $y_t \to
y_t \times (1+3/2 \times c_u)$, essentially rescaling
the overall rate of the $tth$ process. As discussed in the previous section, the
$\ope{G}$ operator induces an additional contribution to the effective
Higgs-gluon coupling, $g_{ggh} \to g_{ggh} (1+ 192 \pi^2/g^2  \times c_G)$, and
thus changes the kinematic distributions. Finally, the $\ope{uG}$ operator
corresponds to a top-quark chromo-dipole moment, which modifies the $gtt$
vertex. It also induces new effective $ggtt$, $gtth$, and $ggtth$ couplings,
promising new kinematic features.

As theory parameters we define the vector
\be
  \theta= (\theta_1,\theta_2,\theta_3)^T = (c_u, 100\,c_G, 100\,c_{uG} )^T \,,
  \label{eq:parameter_space}
\ee
Two of the Wilson coefficients are rescaled by a factor 100 to make sure that
typical values of the three parameters are of the same size. Like in most EFT
analyses, the squared matrix element factorizes as described in
Eq.~\ref{eq:morphing}, and we can use a morphing technique to interpolate event
weights and cross sections from nine benchmarks (or morphing basis points) to
any point in parameter space.

Based on a sample of $1.25\cdot 10^6$ events, we train a
likelihood ratio estimator with the \alices technique and a score estimator with
the \sally method.

\begin{figure*}
\centering
\includegraphics[width=0.49\textwidth]{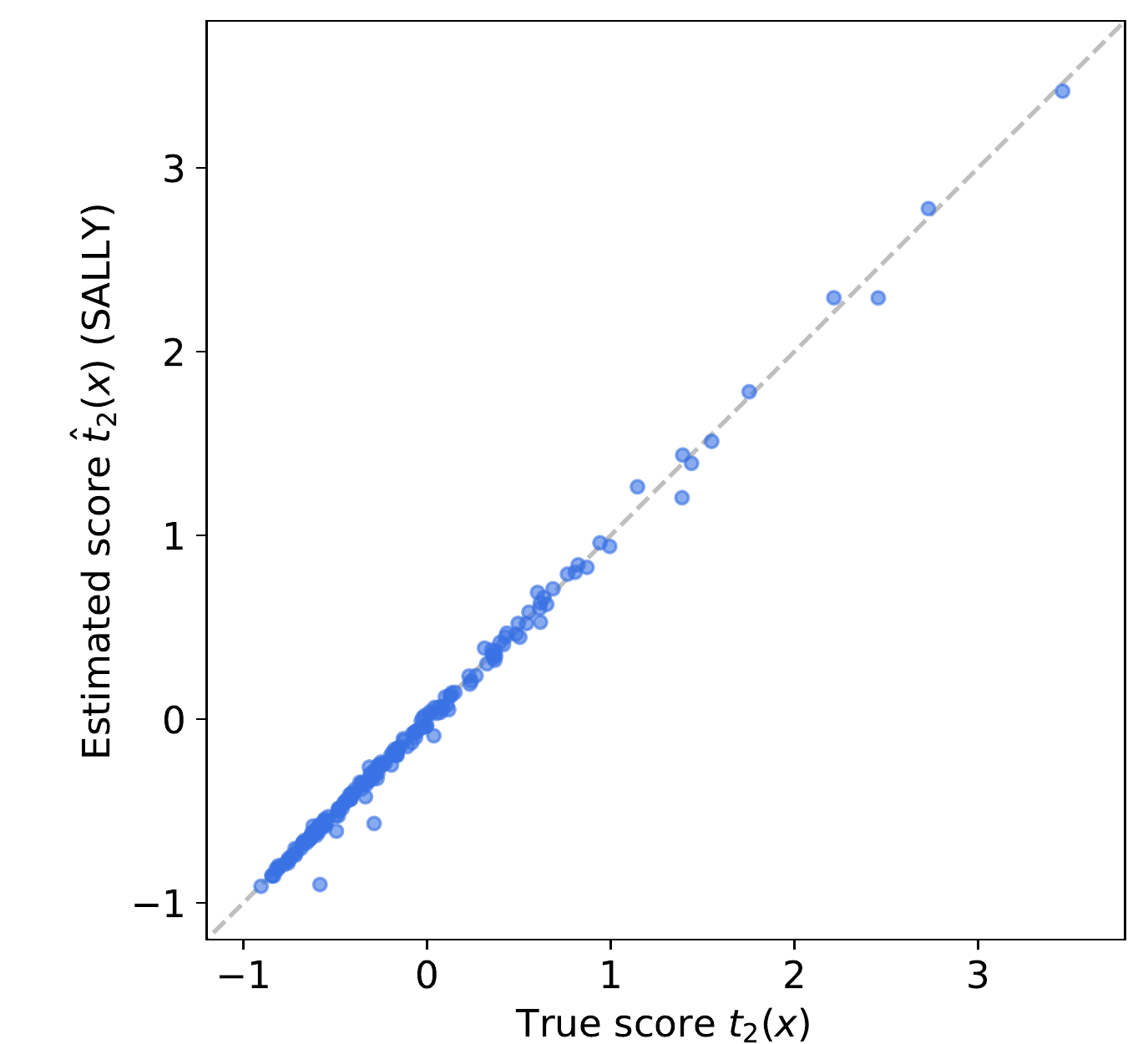}%
\includegraphics[width=0.49\textwidth]{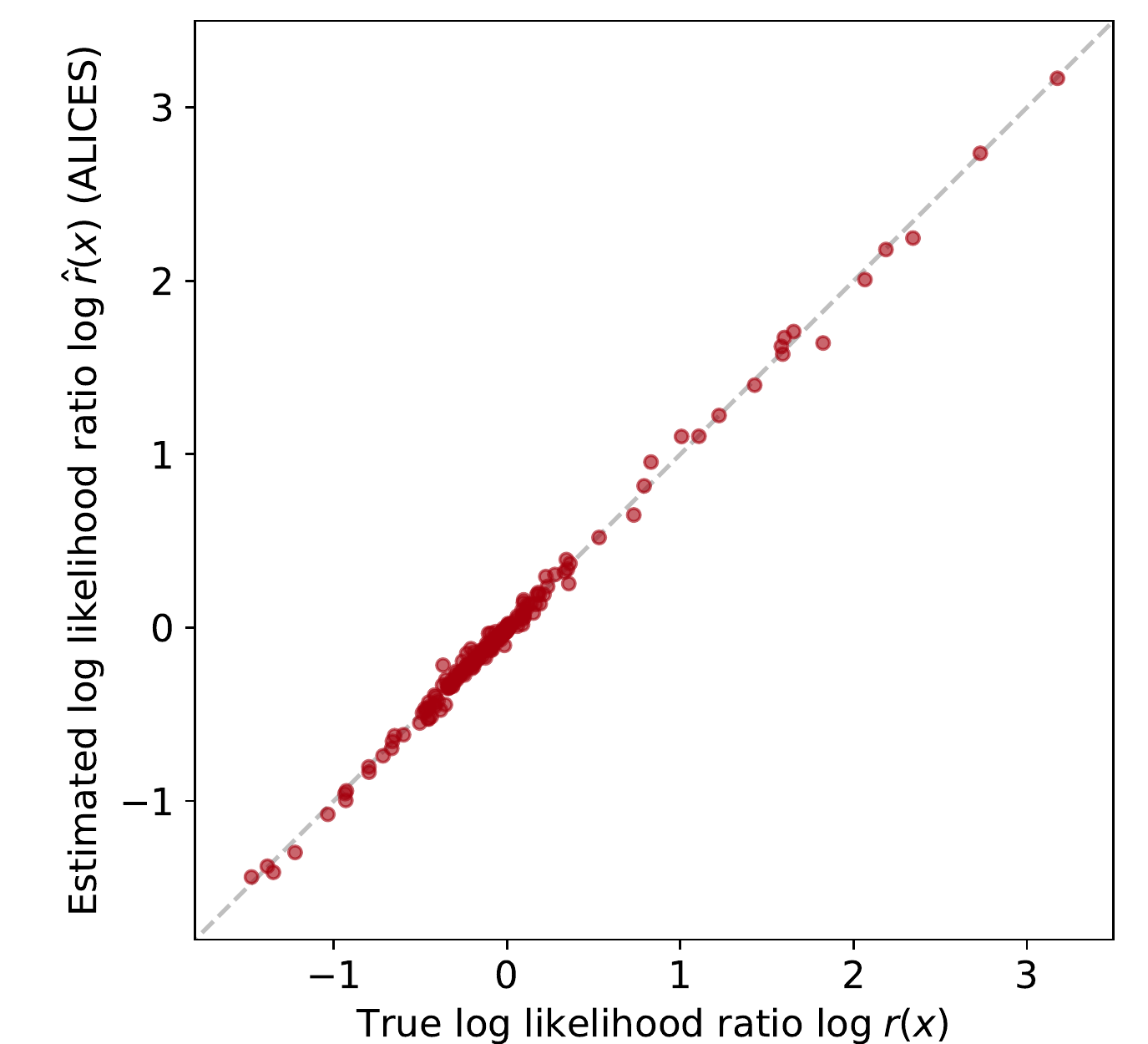}%
\caption{Validation of the analysis techniques in a parton-level analysis,
treating the momenta and flavours of all initial-state and final-state partons
are observable.
\textbf{Left:} Validation of score estimation with the \sally method. Estimated
vs.\ true score component $t_2(x)$ evaluated at the SM.
\textbf{Right:} Validation of likelihood ratio estimation with the \alices
technique. Estimated vs.\ true log likelihood ratio $\log r(x|\theta)$.
The numerator parameter points $\theta$ are drawn from a multivariate
Gaussian with mean $(0, 0, 0)$ and covariance matrix $\diag(0.2^2, 0.2^2,
0.2^2)$ as an example for a relevant region of parameter space.}
\label{fig:validation}
\end{figure*}

We show the results in \figref{validation}. The left panel shows the correlation
between the true and estimated score based on the \sally technique, focusing on
the score component $t_2(x)$ that corresponds to the theory parameter $\theta_2
=100\,c_G$ (with similar results for the other components).
In the right panel we compare the estimated likelihood ratio based
on the \alices method to the ground truth, with parameter points drawn
from a region of parameter space that could be of interest in a typical analysis.
In both cases we find that the
predictions of the neural network are very close to the true values, confirming
that the \madminer{} algorithms work correctly in this truth-level scenario.

\subsection{Realistic physics analysis}
\label{sec:example-results}

Finally we analyse the new physics reach of the $tth$ process in a realistic
setup with high-dimensional event data and theory parameters. We consider
the three dimension-six operators given in Eqs.~\eqref{eq:lagrangian} and
\eqref{eq:operators} and define the theory parameter space as in
Eq.~\eqref{eq:parameter_space}.

In addition to the $tth$ signal we again include the dominant background,
continuum $t\bar{t} \, \gamma\gamma$ production with leptonically decaying tops.
We take into account that this process is sensitive to the theory
parameter $c_{uG}$ through the modified $gtt$ and $ggtt$ vertex while being
independent of $c_{u}$ and $c_{G}$. We neglect subleading backgrounds, in
particular those with fake photons or fake leptons.

The event generation follows the discussion in
Sec.~\ref{sec:example-illustration}; we simulate the parton shower with
\toolfont{Pythia~8} and the detector response with \toolfont{Delphes~3}
using the HL-LHC detector setup. We now
also take into account PDF and scale uncertainties, using the 30 eigenvectors of
the \toolfont{PDF4LHC15\_nlo\_30} PDF set and independently varying
renormalization and factorization scales by a factor of 2.

The event data is described by 48 observables, which includes the
four-momenta of all reconstructed final-state objects (photons, leptons, and jets),
the missing energy, as
well as derived quantities such as the reconstructed transverse momentum of the
di-photon system $p_{T,\gamma\gamma}$. We require the events to pass a di-photon
mass cut $115~\gev<m_{\gamma\gamma}<135~\gev$ and to pass one of four triggers,
which were adopted from the \toolfont{Delphes} default trigger card: the
mono-photon trigger ($p_{T,\gamma}>80~\gev$), the di-photon trigger
($p_{T,\gamma_1}>40~\gev$ and $p_{T,\gamma_2}>20~\gev$), the mono-lepton tigger
($p_{T,\ell}>29~\gev$), or the di-lepton trigger ($p_{T,\ell_1}>17~\gev$ and
$p_{T,\ell_2}>17~\gev$). For an anticipated integrated luminosity of
$\mathcal{L}=3~\iab$ at the HL-LHC, we expect 24.5 $tth$ SM signal and 33.6
$tt\gamma\gamma$ background events to pass these acceptance and selection cuts.

We simulate $1.5\cdot 10^6$ signal and $10^6$ background events (after all cuts)
and extract training samples with $10^7$ unweighted
events. We then train neural networks to estimate the score or likelihood ratio
by minimizing the \sally and \alices loss functions, the latter with a
hyperparameter $\alpha = 0.1$. We use fully connected neural networks with three
hidden layers of 100 units and $\tanh$ activation functions, minimize the loss
functions with the \toolfont{Adam} optimizer using 50
epochs, a batch size of 128, a learning rate that decays exponentially from
$10^{-3}$ to $10^{-5}$, and early stopping to avoid overtraining. These hyperparameters
are the result of a coarse hyperparameter scan, though we did not perform
a exhaustive optimization.

In the final step, we calculate expected exclusion limits and Fisher information
matrices. We compare the results of the new methods to a baseline histogram
analysis of the transverse momentum of the di-photon system
$p_{T,\gamma\gamma}$, and to an analysis of the total cross section alone.

\subsubsection{Fisher information}

\begin{figure*}[t]
\centering
\includegraphics[width=0.49\textwidth]{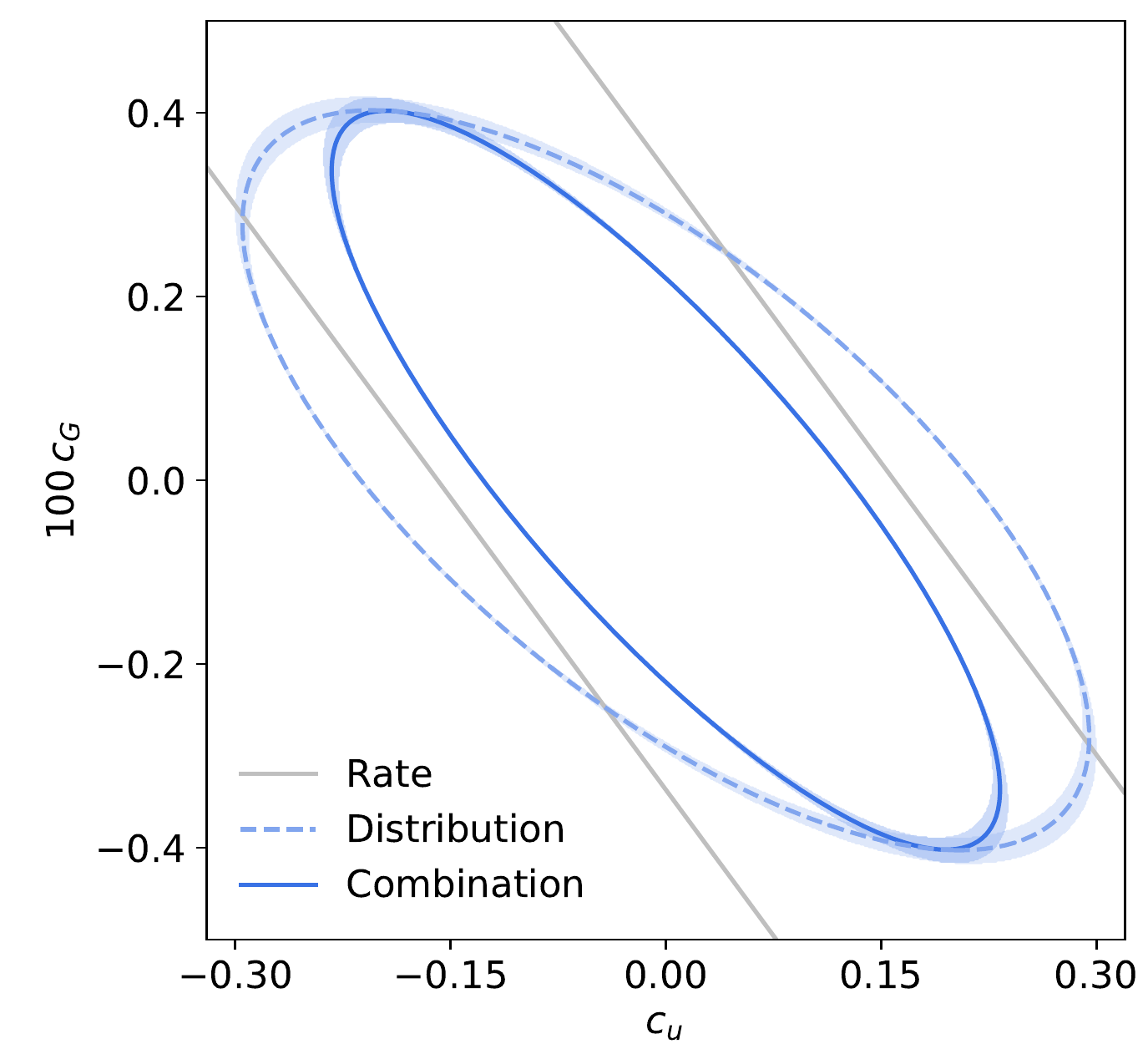}%
\includegraphics[width=0.49\textwidth]{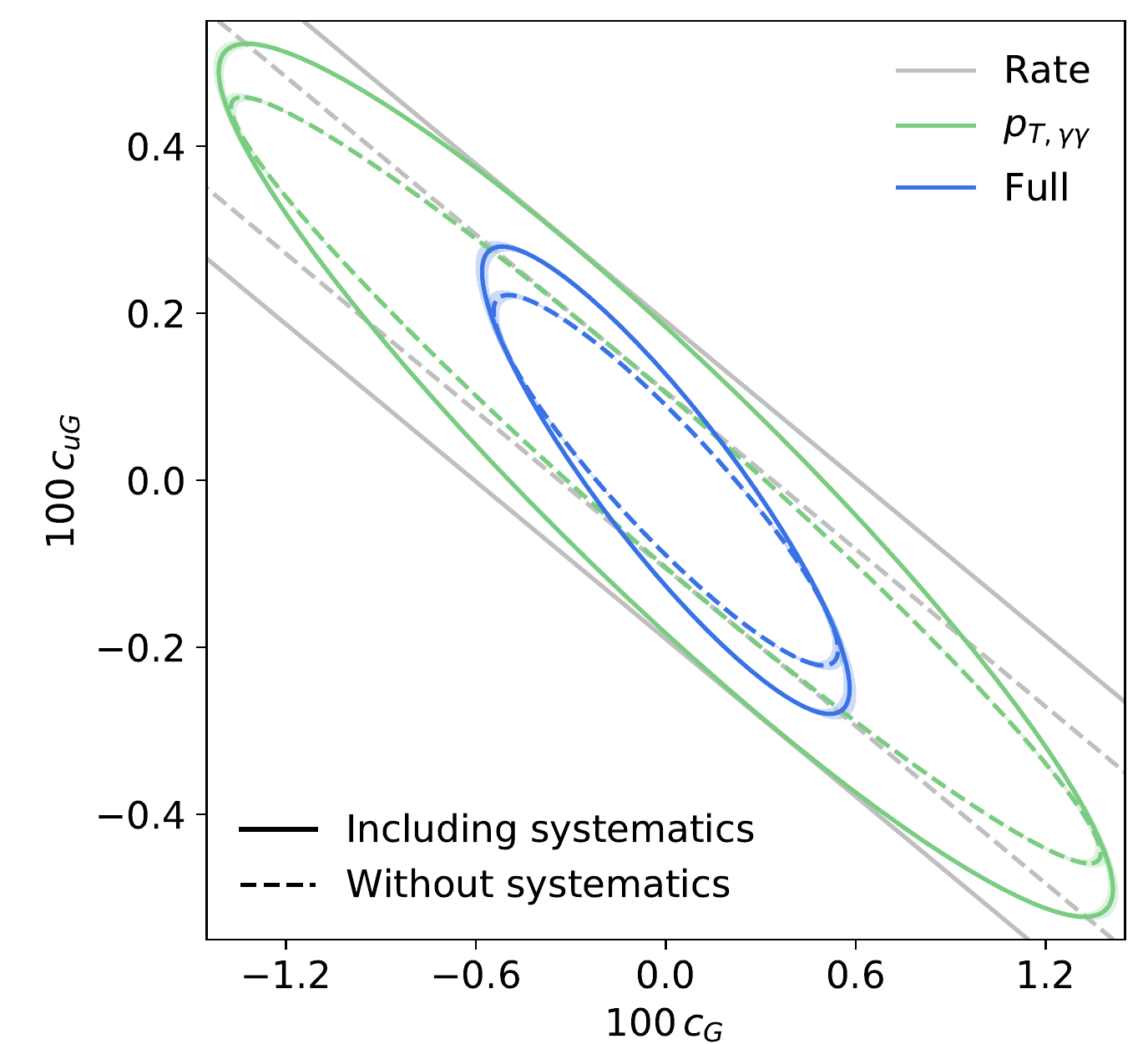}%
\caption{Realistic physics analysis.
\textbf{Left:} Expected $68\%$~CL limits in the $c_G$--$c_u$ plane based on the
Fisher information in the rate (gray), the kinematic information (dashed blue),
and their combination (solid blue). The kinematic information is calculated
based on the \sally technique.  The shaded error bands show the ensemble
variance of a set of 10 independently trained neural networks. We set $c_{uG}$
to zero.
\textbf{Right:} Expected $68\%$~CL limits in the $c_{uG}$--$c_G$ plane based on
the Fisher Information for the rate (grey), a $\ptaa$ histogram
(green), and the full multivariate information based on \sally (blue). The
dashed (solid) line shows the reach without (with) systematic uncertainties. The
shaded error bands show the ensemble variance of a set of 10 independently
trained neural networks. $c_{u}$ is set to zero.}
\label{fig:info_limits}
\end{figure*}

Following our recommendations from \secref{recommendations}, we start our
physics analysis by using the \sally technique, training a neural network to
estimate the score at the SM. We then use it to calculate the SM Fisher
information $I_{ij}$ as described in Sec.~\ref{sec:fisher_info}, finding
\be
  I_{ij} = \threematr {140.5}{68.1}{170.6} {68.1}{47.1}{105.7} {179.5}{105.7}{283.3} \,.
  \label{eq:fisher_info_result}
\ee
This simple matrix summarizes the sensitivity of the measurement on all three
operators. In particular, it allows us to calculate the squared Fisher distance
$d^2(\theta,\thetaref)=I_{ij}(\thetaref) (\theta-\thetaref)_i
(\theta-\thetaref)_j$. As long as $\theta$ is sufficiently close to $\thetaref$,
$d^2$ approximates to $(-2)$ times the expected log likelihood ratio between
$\theta$ and $\thetaref$. That, in turn, can be directly translated into an
expected $p$-value with which $\theta$ can be excluded if $\thetaref$ is true,
using the asymptotic properties of the likelihood ratio~\cite{Wilks:1938dza,
Wald, Cowan:2010js}.  In the following we use the Fisher Information to
calculate expected limits on a combination of two theory parameters, while
fixing the remaining theory parameter to its SM value. In this case, the 68\%
confidence level contours correspond to a local Fisher distance $d=1.509$ (95\%
CL corresponds to $d=2.447$, 99\% CL to $d=3.034$). We show the resulting
expected $68\%$~CL contours in the $c_G$--$c_u$ plane as solid blue line in the
left panel of \figref{info_limits}.

The Fisher information formalism makes it easy to disect these results a little.
First, Eq.~\eqref{eq:fisher_info_lhc} shows that we can separate the full Fisher
information into a rate term and kinematic information. We show this separation
in the left panel of \figref{info_limits} by separately plotting the expected
limits corresponding to the rate information (gray), the kinematic information
(dashed blue) and their combination (solid blue). We find that kinematic
information is crucial for this channel. Since a rate measurement only provides
a single number, at the level of the Fisher information it can only constrain
one direction in theory space and is blind in the remaining direction. This
degeneracy is broken once additional information from the kinematic
distributions is included. Indeed, the kinematic information can constrain the
rate-sensitive direction in theory space almost as well the rate itself.

Another aspect that can be conveniently discussed in the Fisher information
framework are systematic uncertainties. \madminer{} can take PDF and scale
uncertainties into account by parameterizing them with nuisance parameters and
then profiling over them, which at the level of the Fisher information is a
simple matrix operation~\cite{Brehmer:2016nyr, Edwards:2017mnf}. In the right
panel of \figref{info_limits} we analyze the impact of these uncertainties. The
dashed lines show the expected limits neglecting systematic uncertainties, while
the solid lines show results that take systematics into account by profiling
over nuisance parameters. We also again distinguish between the Fisher
information in the rate (gray), the Fisher information in a $\ptaa$ histogram
(green), and the full information based on a neural score estimator (blue). We
can see that the presence of systematic uncertainties, which are dominated by
the scale uncertainty, mainly reduces the sensitivity in the rate-sensitive
direction. The effect of systematic uncertainties is more pronounced for the
information in the total rate and in the $\ptaa$ histogram. The full,
multivariate information is reduced mostly in the rate-sensitive direction in parameter space,
while the information in the orthogonal direction (to which the rate analysis is blind)
is affected only slightly.

The results in both panels of Fig.~\ref{fig:info_limits} do not just include
central predictions for each Fisher information or contour, but also shaded
error bands. These bands visualize the variance of an ensemble of 10 score
estimator instances, each trained on resampled training samples with independent
random seeds. The bands show $2\sigma$ variations, where $\sigma$ is the
ensemble standard deviation for a prediction. The small width of these bands
signals a passed sanity check; a larger width would be an indicator for
numerical issues during training or insufficient training data.

\begin{figure*}[t]
\centering
\includegraphics[width=0.49\textwidth]{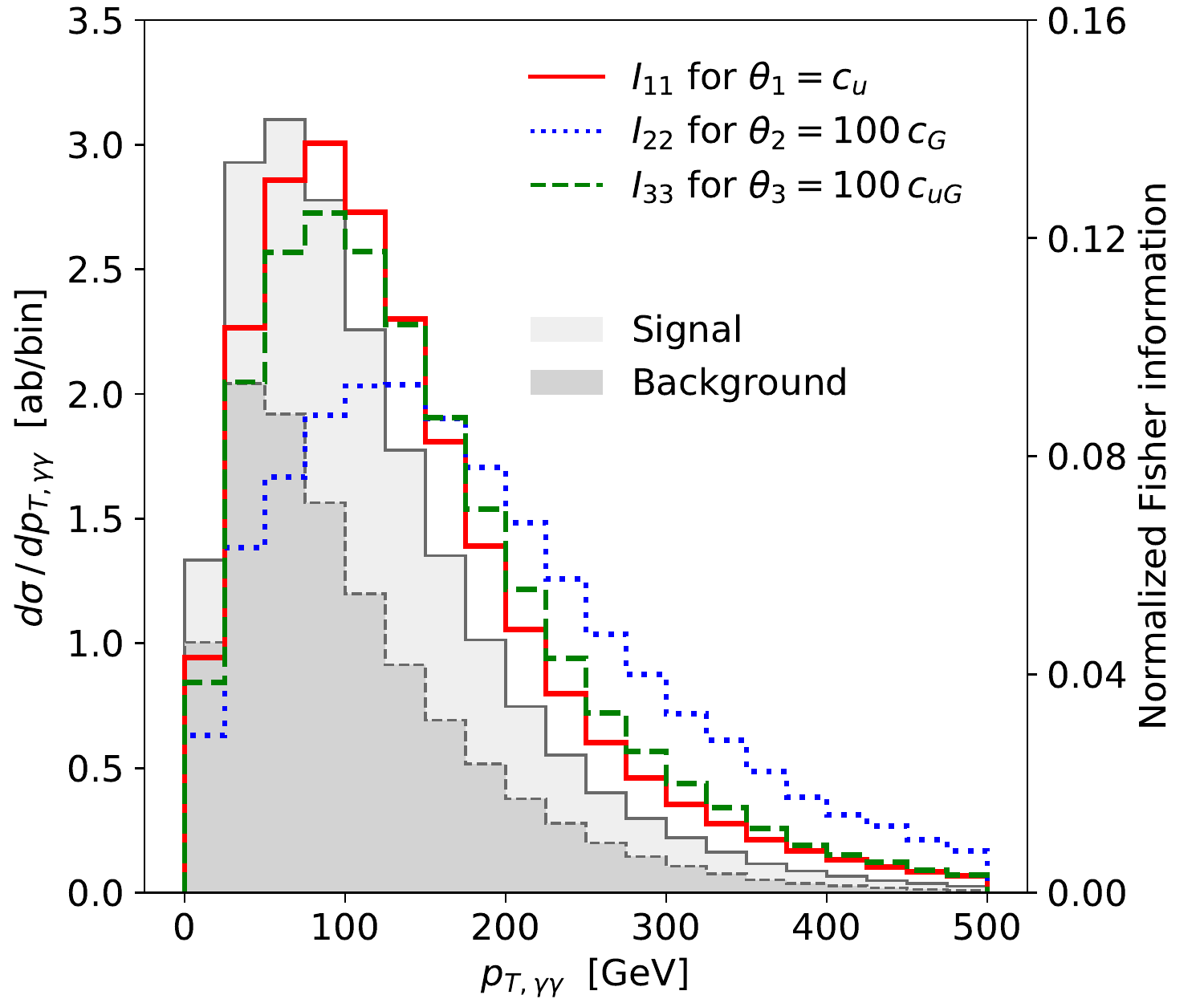}
\includegraphics[width=0.49\textwidth]{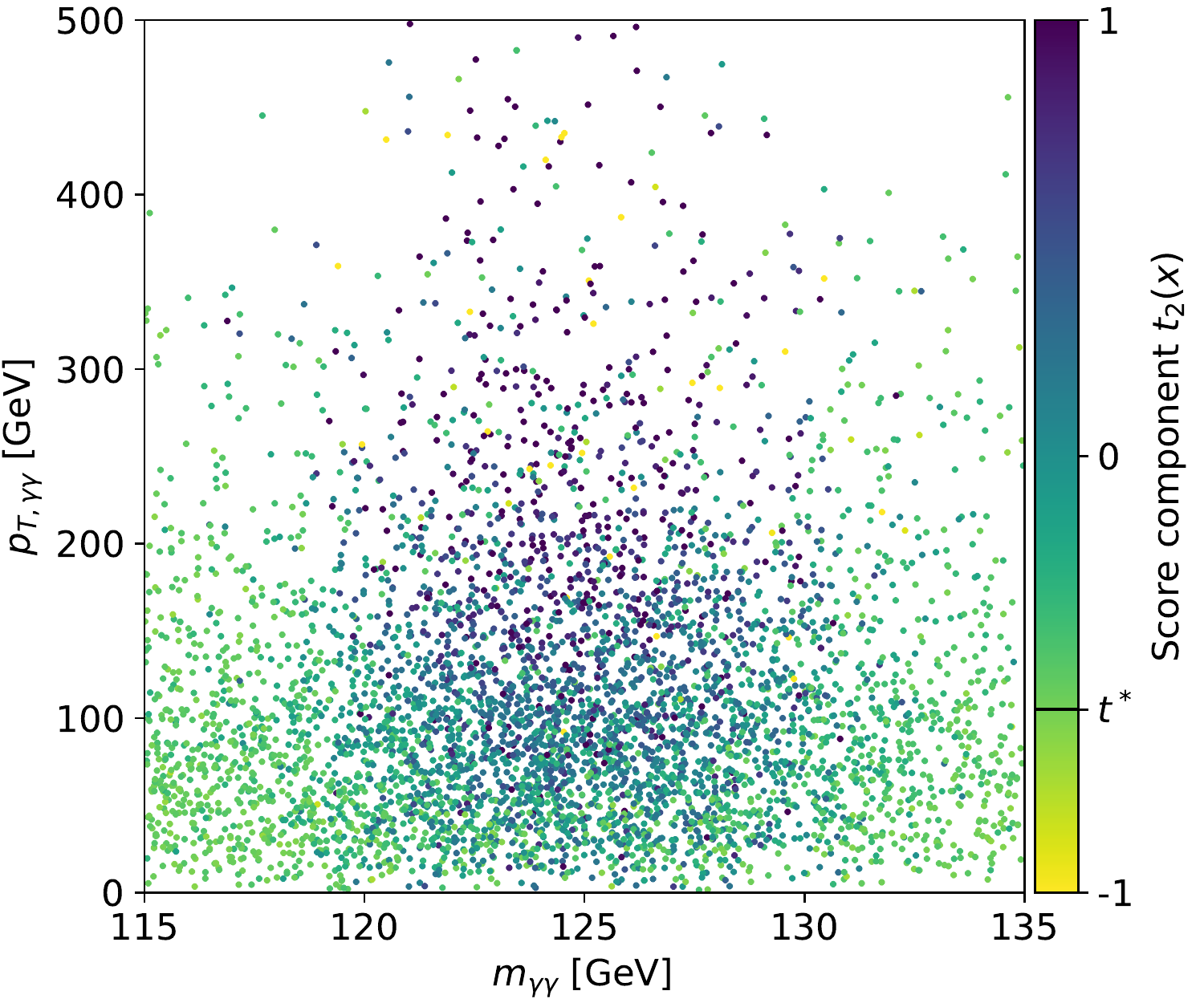}
\caption{Realistic physics analysis.
\textbf{Left:} Differential cross section (shaded grey) and distribution of the
Fisher information components (lines) over $\ptaa$.
\textbf{Right:} Score component $t_2(x)$ corresponding to the Wilson coefficient
$c_{G}$ as a function of di-photon mass $\maa$ and di-photon transverse momentum
$\ptaa$. Note that events in background-dominated regions cluster at the value
$t^* = -0.58$, as discussed in Sec.~\ref{sec:example-illustration}.}
\label{fig:kinematics_vs_info}
\end{figure*}

In the discussion so far we have focused on the total Fisher information
integrated over phase space, which is related to the expected exclusion limits.
There is another useful aspect of the Fisher information: we can analyse the
kinematic distribution of the Fisher information over kinematic variables to
identify the important phase-space regions for a
measurement~\cite{Brehmer:2016nyr}. This knowledge can then be used to design
and optimize the event selection.\footnote{Similarly, important phase-space
regions can also be identified using the log likelihood ratio
directly~\cite{Plehn:2013paa, Kling:2016lay, Goncalves:2018yva}.} As an example we
consider the distribution of information over the di-photon transverse momentum
$\ptaa$, which is shown in the left panel of \figref{kinematics_vs_info}. The
shaded grey areas show the differential cross section for the $tt\gamma\gamma$
background and the SM $tth$ signal. The three colored lines show the normalized
distribution of the diagonal elements of the Fisher Information. We find that
the information on $\ope{u}$, the operator that just rescales the overall $tth$
rate, peaks at $100~\gev$, marking the optimal compromise between good
signal-to-background ratio and large rate. For $\ope{uG}$ and in particular
$\ope{G}$, the information is shifted further towards the high-energy tail of
the distribution, where the kinematic effects from these operators are large.

In the right panel of \figref{kinematics_vs_info} we illustrate the relation
between the score and kinematic variables and show how the score itself can be
used to identify the most sensitive region of phase space. We show the score
component $t_{2}(x)$, corresponding to the Wilson coefficient $c_G$, as a
function of the di-photon mass $\maa$ and di-photon transverse momentum $\ptaa$.
While the $\maa$ distribution for the signal process does not depend on the
Wilson coefficients, this variable is important in telling apart signal and
background contributions. As discussed in \secref{example-illustration},
background events are generated with a constant joint score $t_2(x,z) = t^* =
-0.58$. This is why in kinematic regions dominated by the background, for
instance away from the Higgs mass peak, the estimated score approaches a
constant value $\hat{t}_2(x) \approx t^* = -0.58$. Clusters of positive
(negative) values of the score component correspond to phase-space regions that
are enhanced (suppressed) when increasing $c_G$. The largest scores are observed
for events around the Higgs peak with high $\ptaa \gtrsim 100~\gev$, showing the
increased sensitivity of this high-energy region to the Wilson coefficient
$c_G$. Note that while the score component is clearly related to the two
variables shown here, it is not a simple function of $\maa$ and $\ptaa$; the
neural network instead learned a non-trivial function of the high-dimensional
observable space.

\subsubsection{Exclusion limits}

\begin{figure*}[t]
\centering
\includegraphics[width=0.49\textwidth]{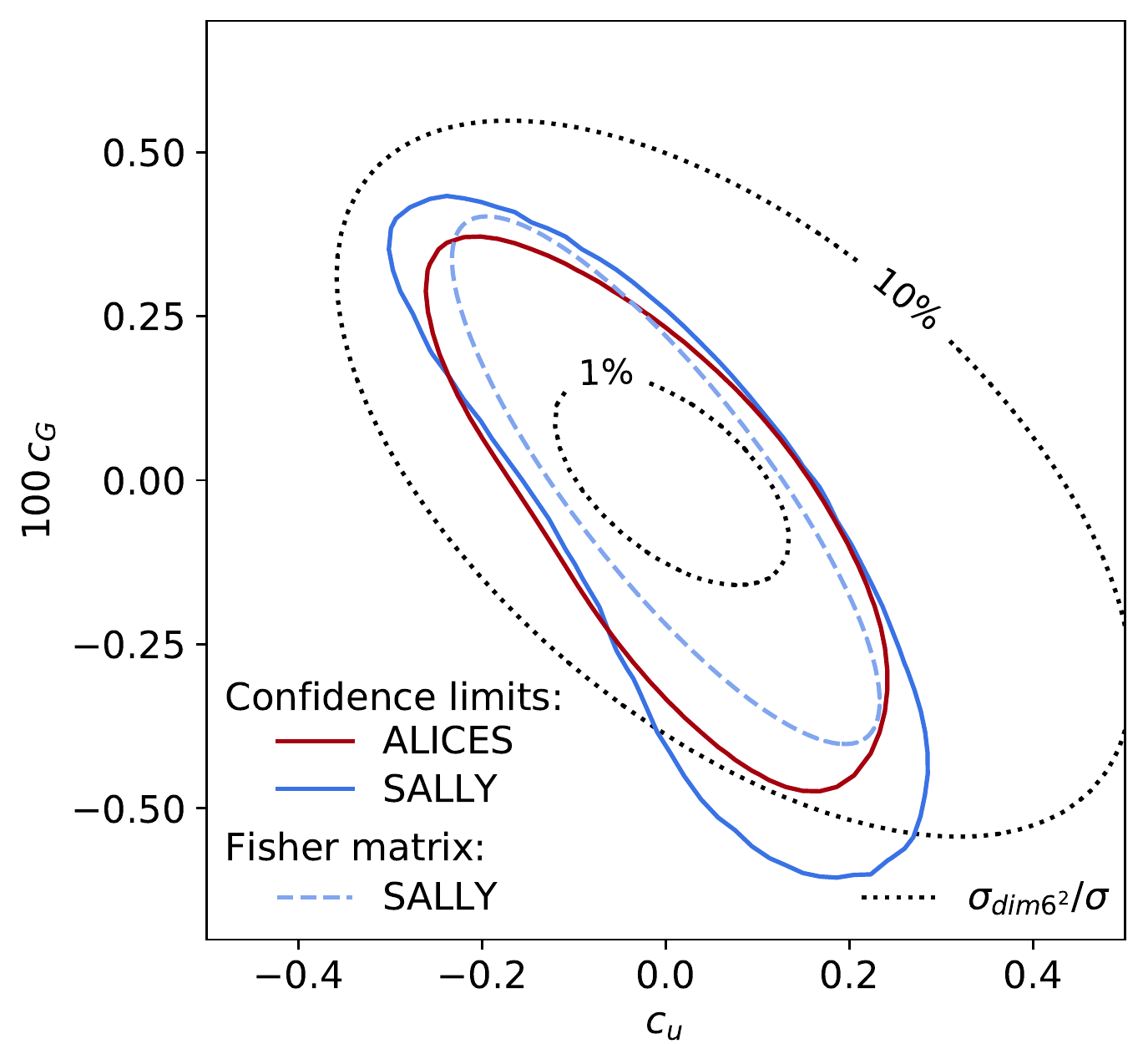}
\includegraphics[width=0.49\textwidth]{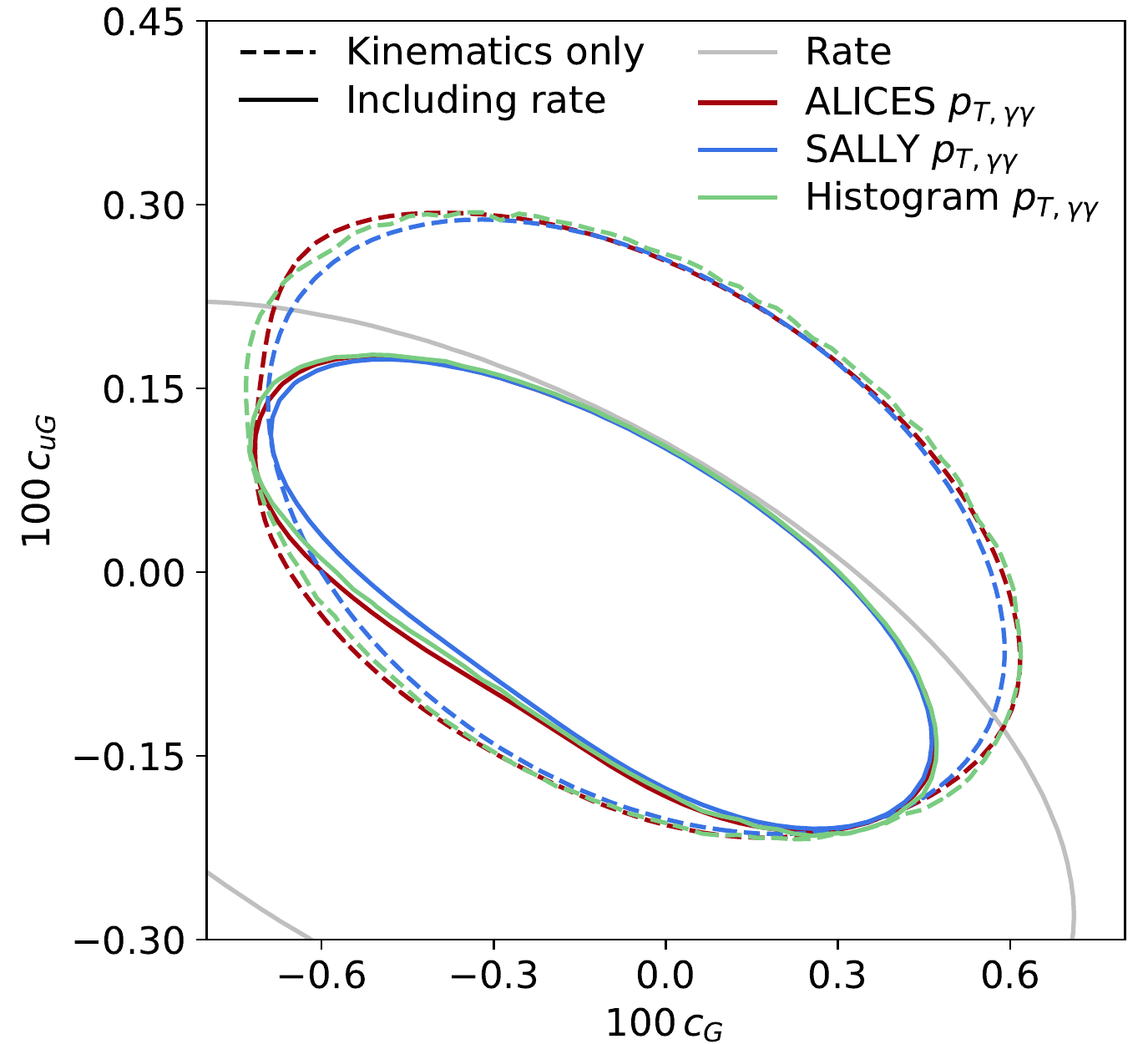}
\caption{Realistic physics analysis.
\textbf{Left:} Comparison of the expected limits in the $c_G$--$c_u$ plane at
68\% CL. We show the limits based on the full likelihood ratio estimated with
the \sally (solid blue) and \alices (solid red) methods as well as approximate
limits based on the Fisher information calculated with \sally (dashed blue). The
dotted black line indicates where the contribution of the dimension-six squared
terms contribute 1\% and 10\% to the total cross section.
\textbf{Right:} Comparison of the expected 68\% CL exclusion limits in the
$c_{uG}$--$c_G$ plane using the rate (gray), a $p_{T,\gamma\gamma}$ histogram
with 20 bins (green), the \sally method trained with only
$p_{T,\gamma\gamma}$ as input (blue), and an \alices likelihood ratio
estimator trained with only $p_{T,\gamma\gamma}$ as input (red). The
dashed limits only use kinematic distributions, while the solid curves include
the rate measurement.
}
\label{fig:limit_checks}
\end{figure*}

So far we have calculated limits in a local approximation, in which non-linear
effects of the theory parameters on the  likelihood function are neglected and
in which the Fisher information fully characterizes the expected log likelihood
ratio as given in \equref{fisher_approx}. Let us now go beyond this
approximation and calculate exclusion limits based on the full likelihood
function, including any non-linear effects. In an analysis of effective
dimension-six operators, the approach in the previous section corresponds to an
analysis of interference effects between the SM contribution and dimension-six
effects, while in this section we also take into account the squared
dimension-six amplitudes. We can thus draw conclusions about the relevance of
the dimension-six squared terms by comparing the limits obtained using the
Fisher information with those obtained using the full likelihood ratio function.

In the left panel of \figref{limit_checks} we show the expected 68\% CL
contours for the parameter plane spanned by $c_G$ and $c_u$. The solid red line
shows the limits obtained using the \alices method, which directly estimates the
likelihood ratio function. The \sally method (solid blue line) estimates the SM
score vector; the components corresponding to $c_G$ and $c_u$ are used as
observables and the likelihood is calculated with two-dimensional histograms.
Finally, the limits based on the local Fisher distance are shown as dashed blue
line. We can see the limits obtained using the three methods do not fully agree,
indicating the relevance of dimension-six squared terms. Indeed, in the region
of parameter space probed at 68\% CL, these terms contribute between 1\% and
10\% to the total rate, as shown by the dotted black lines, and substantially
more in the relevant high-energy region of phase space.

These multivariate results are compared to limits based on the analysis of just
a single summary statistic in the right panel of \figref{limit_checks}. We
analyse the $\ptaa$ distribution with three methods: a histogram with 20 bins
(pastel green), an \alices likelihood ratio estimator trained only on $\ptaa$ as
observable input (forest green), and a \sally estimator of the score trained
only on $\ptaa$ as input (turquoise). We also show limits based only on the
total cross section (grey) and, for each of the three methods, only on kinematic
information (dashed lines). We find that the shape information in the $\ptaa$
distribution (dashed green) is complementary to the rate information, and hence
removes the blind directions of the pure rate measurement. In addition, the
results from the three methods agree very well, providing a non-trivial
cross-check of the three different approaches.

\begin{figure*}[t]
\centering
\includegraphics[width=0.49\textwidth]{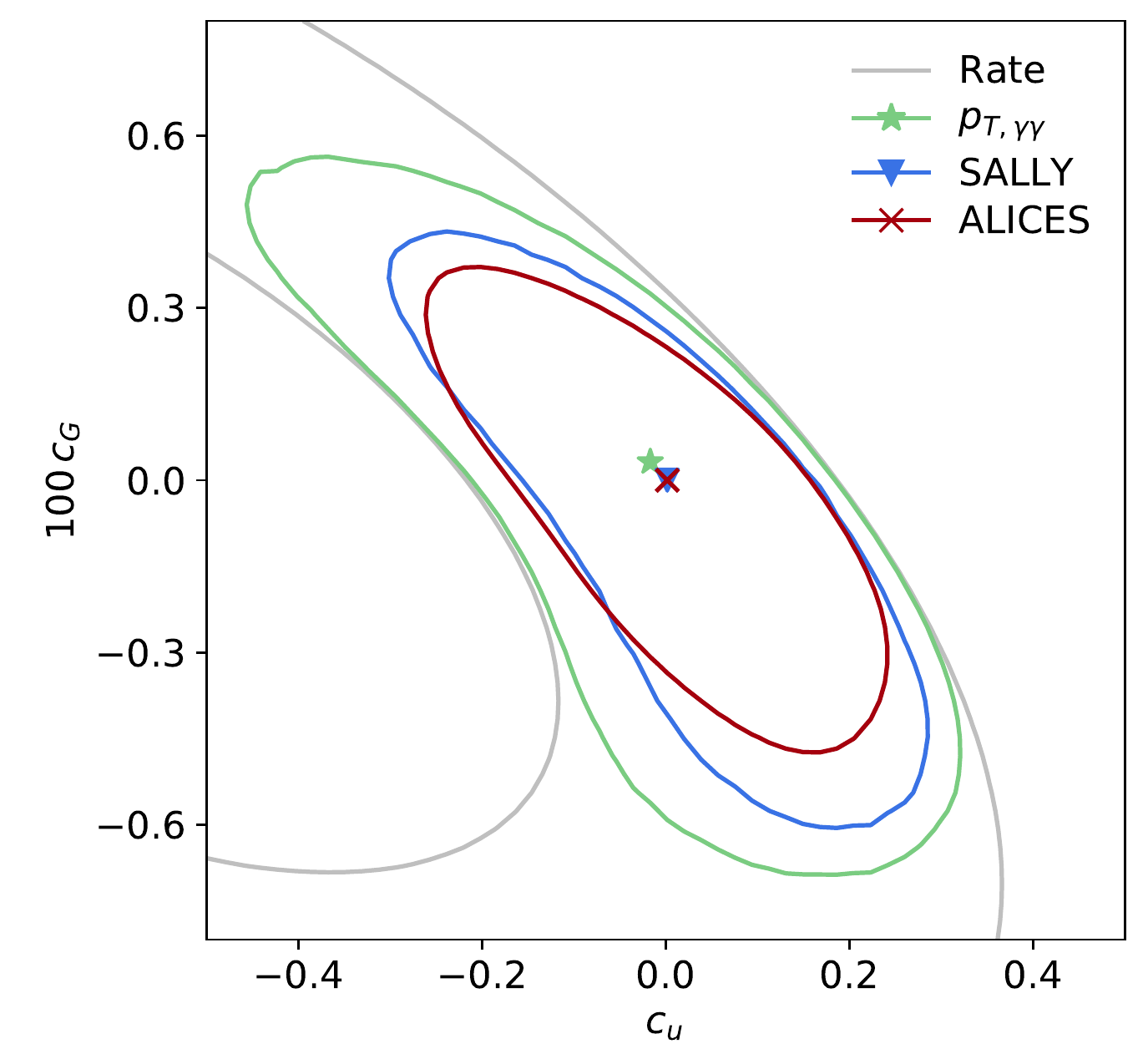}%
\includegraphics[width=0.49\textwidth]{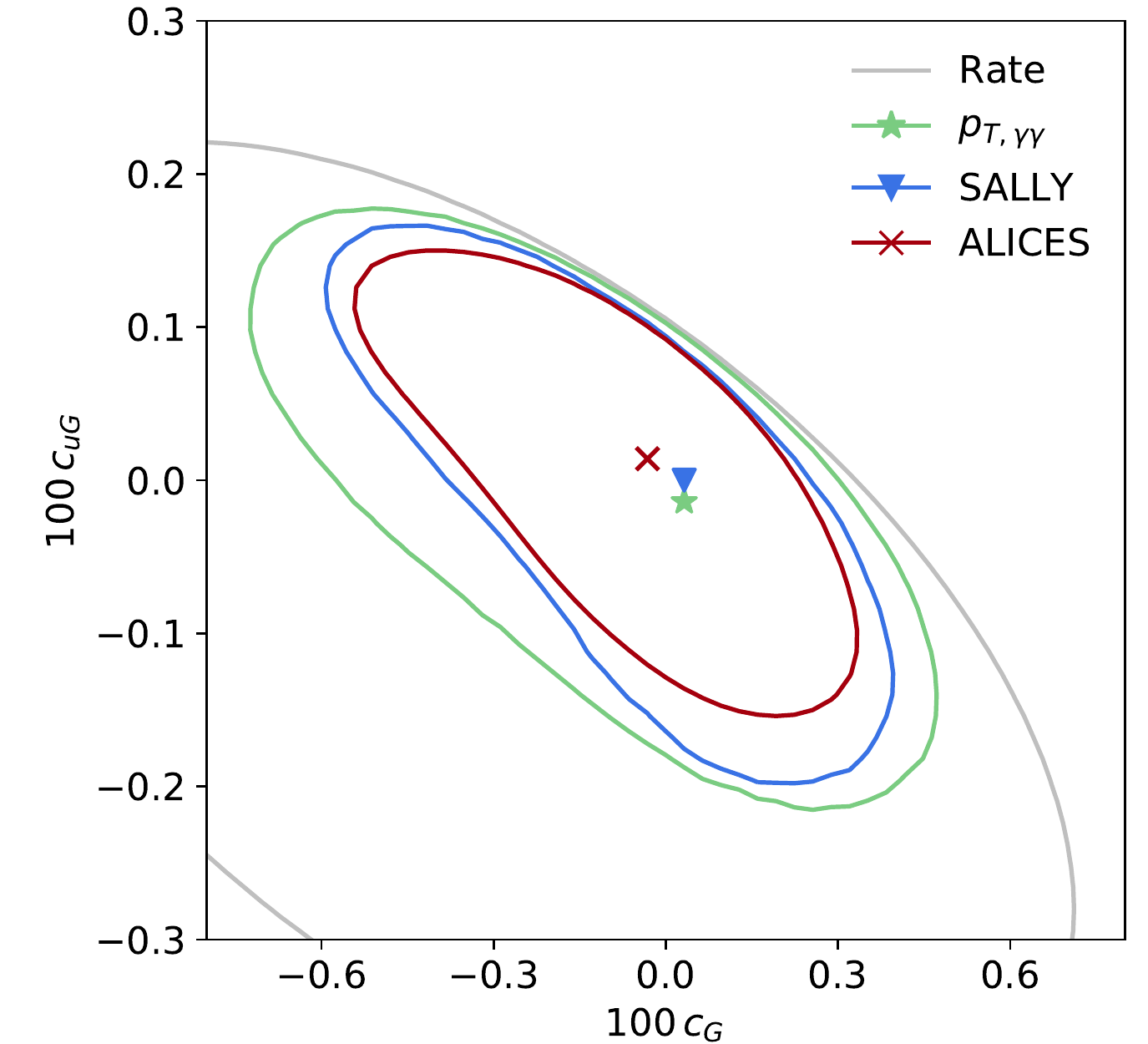}%
\caption{Realistic physics analysis.
Expected exclusion limits on the Wilson coefficients $c_G$ vs.\ $c_u$ with
$c_{uG}$ set to zero (left), and on $c_{uG}$ vs.\ $c_G$ with $c_u$ set to zero
(right). We show best-fit points and $68 \%$ CL limits based on the rate only
(gray), a $p_{T,\gamma\gamma}$ histogram with 20 bins (green), the \sally
technique (blue), and the \alices method (red).}
\label{fig:limit_comparison}
\end{figure*}

Finally we collect the expected limits on the Wilson coefficients based on the
different methods in \figref{limit_comparison}. The left panel shows the
$c_G$--$c_u$ plane, the right panel the  $c_{uG}$--$c_u$ plane, while the
parameter not shown is set to zero in both cases. In grey we show limits based
on a cut-and-count analysis of the total rate. This approach only constrains one
direction in theory space and is blind in the remaining directions. In
particular, this rate-only analysis cannot distinguish between multiple disjoint
best-fit regions, for instance between $c_u = 0$ and $c_u = -4/3$, which
corresponds to a sign-flipped top Yukawa coupling and predicts the same total
cross section.  This degeneracy is broken once kinematic information is
included. Even the simplest case, the histogram-based analysis of a single
variable such as the di-photon transverse momentum $\ptaa$ (green line), can
substantially improve the sensitivity of the analysis.

The blue and red line show the expected limits from the new,
machine-learning-based methods implemented in \madminer{}. In blue we show the
sensitivity of the \sally technique, which uses the estimated score as a vector
of locally optimal observables. We find clearly stronger limits: the score
components are indeed more powerful observables than $\ptaa$. Finally, the red
line shows the limits from the \alices method, in which a neural network learns
the full likelihood ratio function throughout the entire theory parameter space.
In contrast to \sally, it also guarantees optimal sensitivity further away from
the SM reference point, provided that the network was trained successfully\,---\,
and indeed, the \alices technique leads to the strongest expected limits on the
Wilson coefficients.

\section{Conclusions}
\label{sec:conclusions}

In this paper we introduced \madminer{}, a Python package that implements a
range of modern multivariate inference techniques for particle physics
processes. These inference methods require running Monte-Carlo simulations and
extracting additional information related to the matrix elements, using this
information to train neural networks to precisely estimate the likelihood
function, and constraining physics parameters based on this likelihood function
with established statistical methods. \madminer{} implements all steps in this
analysis chain.

These inference techniques are designed for high-dimensional event data without
requiring a choice of low-dimensional summary statistics. Unlike for instance
the matrix element method, they model the effect of a realistic shower and
detector simulation, without requiring any approximations on the underlying
physics. After an upfront training phase, events can be evaluated extremely
fast, which can substantially reduce the computational cost compared to other
methods. Finally, the efficient use of matrix element information reduces the
number of simulated samples required for a succesful training of the neural
networks compared to other, physics-agnostic, machine learning methods.

\madminer{} currently provides interfaces to the simulators
\toolfont{MadGraph5\_aMC}, \toolfont{Pythia~8}, and the fast detector simulation
\toolfont{Delphes~3},  which form a state-of-the-art toolbox for
phenomenological analyses. It supports almost any LHC process, arbitrary theory
models, reducible and irreducible backgrounds, and systematic uncertainties
based on PDF and scale variations. In the future, we are planning to extend
\madminer{} to support detector simulations based on \toolfont{Geant4} as well
as new types of systematic uncertainties.

After discussing the implemented inference techniques and their implementation,
we provided a step-by-step guide through an analysis workflow with \madminer{}.
We then demonstrated the tool in an example analysis of three effective
operators in $tth$ production at the high-luminosity run of the LHC. The
mechanism behind the inference techniques was illustrated in a one-dimensional
case, and the methods validated in a simplified parton-level setup where the
true likelihood is tractable. We demonstrated how \madminer{} lets us isolate
the important phase-space regions and define optimal observables. Finally, we
showed that compared to analyses of the total rate and standard histograms, the
machine-learning-based techniques lead to stronger expected limits on the
effective operators. These results demonstrate that the techniques implemented
in \madminer{} have the potential to clearly improve the sensitivity of the LHC
legacy measurements.


\acknowledgments

We would like to thank Zubair Bhatti, Lukas Heinrich, Alexander Held, and Samuel
Homiller for their important contributions to the development of \madminer{}. We
are grateful to Joakim Olsson for his help with the $tth$ data generation. We
also thank Pablo de Castro, Sally Dawson, Gilles Louppe, Olivier Mattelaer,
Duccio Pappadopulo, Michael Peskin, Tilman Plehn, Josh Rudermann, and Leonora
Vesterbacka for fruitful discussions. Last but not least, we are grateful to the
authors and maintainers of many open-source software packages, including
\toolfont{Delphes~3}~\cite{deFavereau:2013fsa},
\toolfont{Docker}~\cite{Merkel:2014:DLL:2600239.2600241}, \toolfont{Jupyter}
notebooks~\cite{Kluyver2016JupyterN},
\toolfont{MadGraph5\_aMC}~\cite{Alwall:2014hca},
\toolfont{Matplotlib}~\cite{Hunter:2007}, \toolfont{NumPy}~\cite{numpy},
\toolfont{pylhe}~\cite{lukas_2018_1217032},
\toolfont{Pythia~8}~\cite{Sjostrand:2014zea},
\toolfont{Python}~\cite{van1995python},
\toolfont{PyTorch}~\cite{paszke2017automatic},
\toolfont{REANA}~\cite{Simko:2652340},
\toolfont{scikit-hep}~\cite{Rodrigues:2019nct},
\toolfont{scikit-learn}~\cite{scikit-learn},
\toolfont{uproot}~\cite{jim_pivarski_2019_3256257}, and
\toolfont{yadage}~\cite{lukas_heinrich_2017_1001816}.

This work was supported by the U.\,S.~National Science Foundation (NSF) under
the awards ACI-1450310, OAC-1836650, and OAC-1841471. It was also supported
through the NYU IT High Performance Computing resources, services, and staff
expertise. JB and KC are grateful for the support of the Moore-Sloan data
science environment at NYU. KC is also supported through the NSF grant
PHY-1505463, while FK is supported by NSF grant PHY-1620638 and
U.\,S.~Department of Energy grant DE-AC02-76SF00515.

\appendix
\section{Frequently asked questions}
\label{sec:faq}

Here we collect questions that are asked often, hoping to avoid misconceptions:

\begin{itemize}
  \item \emph{Does the whole event history not change when I change parameters?}

  No. In probabilistic processes such as those at the LHC, any given event
  history is typically compatible with different values of the theory
  parameters, but might be more or less likely. With ``event history'' we mean
  the entire evolution of a simulated particle collision, ranging from the
  initial-state and final-state elementary particles through the parton shower
  and detector interactions to observables. The joint likelihood ratio and joint
  score quantify how much more or less likely one particular such evolution of a
  simulated event becomes when the theory parameters are varied.

  \item \emph{If the network is trained on parton-level matrix element
  information, how does it learn about the effect of shower and detector?}

  It is true that the ``labels'' that the networks are trained on, the joint
  likelihood ratio and joint score, are based on parton-level information. But
  the inputs into the neural network are observables based on a full simulation
  chain, after parton shower, detector effects, and the reconstruction of
  observables. It was shown in Ref.~\cite{Brehmer:2018hga, Brehmer:2018kdj,
  Brehmer:2018eca} that the joint likelihood ratio and joint score are unbiased,
  but noisy, estimators of the true likelihood ratio and true score (including
  shower and detector effects). A network trained in the right way will
  therefore learn the effect of shower and detector. We illustrate this
  mechanism in Sec.~\ref{sec:example-illustration} in a one-dimensional problem.

  \item \emph{Can this approach be used for signal-background classification?}

  Yes. In the simplest case, where the signal and background hypothesis do not
  depend on any additional parameters, the \carl, \rolr, or \alice techniques
  can be used to learn the probability of an individual event being signal or
  background. If there are parameters of interest such as a signal strength or
  the mass of a resonance, the score becomes useful and techniques such as
  \sally, \rascal, \cascal, and \alices can be more powerful.

  The techniques that use the joint likelihood ratio or score require less
  training data when the signal and background processes populate the same
  phase-space regions. If this is not the case, these methods still apply, but
  will not offer an advantage over the traditional training of binary
  classifiers.

  \item \emph{What if the simulations do not describe the physics accurately?}

  No simulator is perfect, but many of the techniques used for incorporating
  systematic uncertainties from mismodeling in the case of multivariate
  classifiers can also be used in this setting. For instance, often the effect
  of mismodeling can be corrected with simple scale factors and the residual
  uncertainty incorporated with nuisance parameters. \madminer{} can handle such
  systematic uncertainties as discussed above. If only particular phase-space
  regions are problematic, for instance those with low-energy jets, we recommend
  to exclude these parameter regions with suitable selection cuts. If the
  kinematic distributions are trusted, but the overall normalization is less
  well known, a data-driven normalization can be used.

  Of course, there is no silver bullet, and if the simulation code is not
  trustworthy at all in a particular process and the uncertainty cannot be
  quantified with nuisance parameters, these methods (and many more traditional
  analysis methods) will not provide accurate results.

  \enlargethispage{\baselineskip}

  \item \emph{Is the neural network a black box?}

  Neural networks are often criticized for their lack of explainability. It is
  true that the internal structure of the network is not directly interpretable,
  but in \madminer{} the interpretation of what the network is trying to learn
  is clearly connected to the matrix element. In practical terms, one of the
  challenges is to verify whether a network has been successfully trained. For
  that purpose, many cross-checks and diagnostic tools are available to make
  sure that this is the case:
  \begin{itemize}
    \item checking the loss function on a separate validation sample;
    \item training of multiple network instances with independent random seeds,
    as discussed above;
    \item checking the  expectation values of the score and likelihood ratio
    against their known true values, see Ref.~\cite{Brehmer:2018eca};
    \item varying of the reference hypothesis in the likelihood ratio, see
    Ref.~\cite{Brehmer:2018eca};
    \item training classifiers between data reweighted with the estimated
    likelihood ratio and original data from a new parameter point, see
    Ref.~\cite{Brehmer:2018eca};
    \item validating the inference techniques in low-dimensional problems with
    histograms, see Sec.~\ref{sec:example-illustration};
    \item validating the inference techniques on a parton-level scenario with
    tractable likelihood function, see Sec.~\ref{sec:example-validation}; and
    \item checking the asymptotic distribution of the likelihood ratio
    against~Wilks' theorem~\cite{Wilks:1938dza, Wald, Cowan:2010js}.
  \end{itemize}

  Finally, when limits are set based on the Neyman construction with toy
  experiments (rather than using the asymptotic properties of the likelihood
  ratio), there is a coverage guarantee: the exclusion contours constructed in
  this way will not exclude the true point more often than the confidence level.
  No matter how wrong the likelihood, likelihood ratio, or score function
  estimated by the neural network is, the final limits might lose statistical
  power, but will never be too optimistic.

  \item \emph{Are you trying to replace PhD students with a machine?}

  As a preemptive safety measure against scientists being made redundant by
  automated inference algorithms, we have implemented a number of bugs in
  \madminer{}. It will take skilled physicists to find them, ensuring safe jobs
  for a while. More seriously, just as \toolfont{MadGraph} automated the process
  of generating events for an arbitrary hard scattering process, \madminer{}
  aims to contribute to the automation of several steps in the inference chain.
  Both developments enhance the productivity of physicists.
\end{itemize}


\bibliography{references}

\end{document}